\documentclass[11pt]{article}

\usepackage{a4wide}
\usepackage{amsmath}
\usepackage{amssymb}
\usepackage{amsfonts}
\usepackage{amsthm}
\usepackage{array}
\usepackage{xy}
\usepackage{enumerate}
\usepackage{graphicx}
\usepackage{float}

\newcolumntype{C}[1]{>{\centering\arraybackslash$}m{#1}<{$}}
\newlength{\mycolwd}

\begin{document}

\begin{center}


{\LARGE{Visualizing extremal positive maps in unital and trace preserving form}}

\bigskip
\bigskip

{\large{Leif Ove Hansen and Jan Myrheim\\
Department of Physics, Norwegian University of Science and Technology,\\
N-7491 Trondheim, Norway}}

\bigskip
\bigskip

{\large{\today}}

\end{center}

\bigskip

\begin{abstract}
  We define an entanglement witness in a composite quantum system as
  an observable having nonnegative expectation value in every
  separable state.  Then a state is entangled if and only if it has a
  negative expectation value of some entanglement witness.  Equivalent
  representations of entanglement witnesses are as nonnegative
  biquadratic forms or as positive linear maps of Hermitian matrices.
  As reported elsewhere, we have studied extremal entanglement
  witnesses in dimension $3\times 3$ by constructing numerical
  examples of generic extremal nonnegative forms.  These are so
  complicated that we do not know how to handle them other than by
  numerical methods.  However, the corresponding extremal positive
  maps can be presented graphically, as we attempt to do in the
  present paper.  We understand that a positive map is extremal when
  the image of $\mathcal{D}$, the set of density matrices, fills out
  $\mathcal{D}$ maximally, in a certain sense.  For the graphical
  presentation of a map we transform it to a standard form where it is
  unital and trace preserving.  We present an iterative algorithm for
  the transformation, which converges rapidly in all our numerical
  examples and presumably works for any positive map.  This standard
  form of an entanglement witness is unique up to unitary product
  transformations.
\end{abstract}


\section{Introduction}

The phenomenon of quantum entanglement \cite{Schrodinger,EPR1935} is
central to the development and understanding of quantum information
theory and computation \cite{Nielsen}. For a general bipartite mixed
quantum state it is not known how to determine in an efficient way
whether the state is entangled or separable.  An experimental setup
usually involves mixed quantum states, and the need to control,
manipulate and quantify entanglement in such states is therefore of
fundamental importance.  The characterization of entangled states of a
quantum system composed of two subsystems thus remains one of the key
unsolved problems in the theory of quantum information.  We will
consider here only the finite dimensional case.

Entanglement of a quantum state may be revealed experimentally through
negative expectation values of certain observables, called
entanglement witnesses, that have positive expectation values in all
separable states~\cite{MPRHorodecki96,Barbieri2003}.  The
Choi--Jamio{\l}kowski isomorphism relates the entanglement witnesses
to positive maps on matrix
algebras~\cite{Jamiolkowski1972,Choi1975}. In this way the study of
entanglement witnesses is closely related to the study of positive
maps.  This study was pioneered by St{\o}rmer~\cite{Stormer1963}, who
has recently given an extensive review~\cite{Stormer2013}.

In order to investigate the separability problem using positive maps
we need to understand the structure of the set of positive maps.
Unfortunately, this is a problem which seems just as difficult as the
original separability problem.  Since the set of positive maps is a
compact convex set, a classification of the extremal positive maps
would entail a classification of all positive maps.  This is our
motivation for studying the extremal positive maps.


\subsection*{Outline of this paper}

The contents of the paper are organized in the following manner.

In section 2 we review some basic theory regarding positive maps and
entanglement witnesses.  We define the link between entanglement
witnesses and positive maps, known as the Choi--Jamio{\l}kowski
isomorphism.  An important class of positive maps, the completely
positive maps, is defined.  These maps are quantum operations, since
they map quantum states into quantum states.  The important
distinction between decomposable and non-decomposable maps is made,
and the relation to the Peres separability criterion~\cite{Peres1996}
is shown.  Furthermore, the unital and trace preserving properties of
positive maps are defined, and the relation between these two
properties is established.  These two concepts are further discussed
in Section~3.  Finally, the expectation values of entanglement
witnesses in pure product states are introduced.  These are positive
semidefinite biquadratic forms, the zeros of which play an important
role.  The importance of the zeros for the understanding of the
extremal entanglement witnesses and the corresponding extremal
positive maps is outlined.

In section 3 we argue that any entanglement witness may be transformed
into a unital and trace preserving form through a product
transformation. We use the fact that a map is trace preserving if and
only if the transposed map is unital to define conditions on this
product transformation. We then suggest an iteration scheme to solve
the resulting equations, and we define and investigate the conditions
under which this iteration scheme will converge.

Finally, in Section~4 we use the unital and trace preserving form of a
generic extremal entanglement witness in the $3\times 3$ system that
we have produced numerically~\cite{Hansen2014}, to create various
plots showing how the corresponding positive map acts.  We present
similar plots illustrating the action of the Choi--Lam extremal
positive map.  We also discuss briefly an example of extremal
witnesses in dimension $2\times 4$.

\section{Preliminary theory}

In this section we introduce our notation and review some mathematical
background.

We write $H_k$ for the set of Hermitian $k\times k$ matrices, which is
a real Hilbert space with the natural scalar product
\begin{equation}
\langle X,Y\rangle=\textrm{Tr}\,(XY)\;.
\end{equation}

The mathematical description of a finite dimensional composite quantum
system involves the complex tensor product
$\mathbb{C}^N=\mathbb{C}^m\otimes\mathbb{C}^n$ and the real tensor
product $H_N=H_m\otimes H_n$ with $N=mn$.
We write the components of $\psi\in\mathbb{C}^N$ as $\psi_I=\psi_{ij}$
where
\begin{equation}
I=1,2,\ldots,N
\leftrightarrow ij=11,12,\ldots,1n,21,\ldots,mn\;.
\end{equation}
The matrix elements of an $N\times N$ matrix $A$ are
$A_{IK}=A_{ij;kl}$ with
$I\leftrightarrow ij$ and $K\leftrightarrow kl$.  A product vector
$\psi=\phi\otimes\chi$ has components $\psi_{ij}=\phi_i\chi_j$, and a
product matrix $A=B\otimes C$ has components $A_{ij;kl}=B_{ik}C_{jl}$.
We define the partial transpose $A^P$ as the transpose with respect to
the second factor in the tensor product,
\begin{equation}
(A^P)_{ij;kl}=A_{il;kj}\;.
\end{equation}
%

We use {\em density matrices}, positive matrices of unit trace, to
represent physical states of the most general kind, so called mixed
states.  The set of all $N\times N$ density matrices we call
${\cal D}_N$, or ${\cal D}$ if we need not specify the dimension.  It
is a compact convex set of dimension $N^2-1$, completely determined by
its extremal points, which are the pure states $\rho=\psi\psi^{\dag}$
with $\psi\in\mathbb{C}^N$.

It is a remarkable fact that the partial transpose $\rho^P$ of a
density matrix $\rho$ need not be positive definite.  A density matrix
with a positive partial transpose is called a PPT state, thus we
define the set of PPT states as
$\mathcal{P}=\mathcal{D}\cap\mathcal{D}^P$.

\subsection{The geometry of density matrices}

A density matrix $\rho\in\mathcal{D}_N$ lies on the boundary
$\partial\mathcal{D}_N$ if it has rank $k<N$.  Then it lies in the
interior of a face $\mathcal{D}_k(\mathcal{U})$, which is the set of
density matrices on a $k$ dimensional subspace
$\mathcal{U}\subset\mathbb{C}^N$.  Every face of $\mathcal{D}_N$ is of
the type $\mathcal{D}_k(\mathcal{U})$.  Thus $\partial\mathcal{D}_N$
consists of faces arranged in the same hierarchical structure as the
lattice of subspaces of $\mathbb{C}^N$.

A $2\times 2$ density matrix has the form
\begin{equation}
\rho=\frac{1}{2}
\begin{pmatrix}
1+z & x-\textrm{i} y \\
x+\textrm{i} y & 1-z
\end{pmatrix}
\end{equation}
with $x,y,z$ real and $x^2+y^2+z^2\leq 1$.  Thus $\mathcal{D}_2$ is a
three dimensional sphere, the Bloch sphere.  The boundary states, with
$x^2+y^2+z^2=1$, are the pure states.

The set $\mathcal{D}_3$ of $3\times 3$ density matrices has dimension
eight.  Its boundary $\partial\mathcal{D}_3$ consists of a seven
dimensional set of rank two matrices, and a four dimensional set of
rank one matrices, the pure states.  Every rank two state in
$\mathcal{D}_3$ lies in the interior of a Bloch sphere
$\mathcal{D}_2(\mathcal{U})$.

\subsection{Entanglement and entanglement witnesses}

A density matrix $\rho\in\mathcal{D}_N$ is {\em separable} if it is of
the form
\begin{equation}
\label{eq:sepdef}
\rho=\sum_ap_a\,\sigma_a\otimes\tau_a\;,
\end{equation}
where $\sigma_a\in\mathcal{D}_m$, $\tau_a\in\mathcal{D}_n$, and
$p_a>0$, $\sum_ap_a=1$.  Its partial transpose is then positive.  This
fact, that $\mathcal{S}\subset\mathcal{P}$ where $\mathcal{S}$ is the
set of separable states, provides an easy test for separability, known
as the Peres criterion~\cite{Peres1996}.  A density matrix is {\em
  entangled} if it is not separable.  The separability problem, how to
decide whether a given density matrix is separable or entangled, is
difficult because it is difficult to recognize the entangled PPT
states.

We define the dual set
\begin{equation}
{\cal S}^{\circ}
=\{\,A\in H_N\mid\textrm{Tr}(A\rho)\geq 0\quad
\forall\rho\in{\cal S}\,\}\;.
\end{equation}
The dual of ${\cal S}^{\circ}$ is ${\cal S}$, thus $\rho$ is separable
if and only if $\textrm{Tr}(A\rho)\geq 0$ for every
$A\in{\cal S}^{\circ}$.  For this reason we call a matrix
$A\in{\cal S}^{\circ}$ an {\em entanglement witness}.  Note that we do
not impose here the usual condition on an entanglement witness $A$
that it should have at least one negative eigenvalue, so that
$\textrm{Tr}(A\rho)<0$ for some $\rho\in\mathcal{D}_N$.

The convex set ${\cal S}$ is completely determined by its extremal
points, which are the pure product states
$\rho=\psi\psi^{\dag}=(\phi\phi^{\dag})\otimes(\chi\chi^{\dag})$ with
$\psi=\phi\otimes\chi$ and $\phi^{\dag}\phi=\chi^{\dag}\chi=1$, for
which
\begin{equation}
\textrm{Tr}(A\rho)
=\psi^{\dag}A\psi
=(\phi\otimes\chi)^{\dag}A(\phi\otimes\chi)
=\sum_{i,j,k,l}\phi_i^{\ast}\chi_j^{\ast}A_{ij;kl}\phi_k\chi_l\;.
\end{equation}
Thus $A\in{\cal S}^{\circ}$ if and only if the biquadratic form
\begin{equation}
\label{eq:biqform}
f_A(\phi,\chi)=(\phi\otimes\chi)^{\dag}A(\phi\otimes\chi)
\end{equation}
is non-negative definite.  Since
\begin{equation}
(\phi\otimes\chi)^{\dag}A^P(\phi\otimes\chi)=
(\phi\otimes\chi^{\ast})^{\dag}A(\phi\otimes\chi^{\ast})
\end{equation}
we see that $A^P\in{\cal S}^{\circ}$ if and only if
$A\in{\cal S}^{\circ}$.


\subsection{Maps}

A {\em map}, in the terminology used here, is a linear transformation
$\textrm{\bf{M}}:H_m\mapsto H_n$ such that $Y=\textrm{\bf{M}}X$ when
\begin{equation}
\label{CJ1}
Y_{jl}=\sum_{i,k}M_{jl;ik}X_{ik}\;.
\end{equation}
The complex matrix elements $M_{jl;ik}$ defining the map satisfy the
relations
$M_{jl;ik}=(M_{lj;ki})^{\ast}$,
%
so that the map preserves Hermiticity.  The relation
\begin{equation}
A_{ij;kl}=M_{jl;ki}
\end{equation}
defines a one to one correspondence between a Hermitian matrix
$A\in H_N$ and a map $\textrm{\bf{M}}=\textrm{\bf{M}}_A$.  As we
define the correspondence here it differs slightly from the well known
Choi--Jamio{\l}kowski isomorphism, which associates the matrix $A$
with the map
%
$X\mapsto\textrm{\bf{M}}_A(X^T)\,$.
%

It is useful to introduce matrices $E_a\in H_m$ and $F_b\in H_n$ that
are basis vectors in the two Hilbert spaces.  Then the map ${\bf{M}}$
is given by real matrix elements $M_{ba}$ such that
\begin{equation}
\textrm{\bf{M}}E_a=\sum_b M_{ba}F_b\;.
\end{equation}
We will choose the basis vectors to be orthonormal, then the matrix
elements are
\begin{equation}
M_{ba}=\langle F_b,\textrm{\bf{M}}E_a\rangle\;.
\end{equation}
Furthermore, we will choose the unit matrices as basis vectors,
$E_0={I_m}/{\sqrt{m}}$, $F_0={I_n}/{\sqrt{n}}$.
%
%
Then the matrices $E_a$ for $a=1,2,\ldots,m^2-1$ and $F_b$ for
$b=1,2,\ldots,n^2-1$ are traceless.

The transposed map $\textrm{\bf{N}}=\textrm{\bf{M}}^T:H_n\mapsto H_m$
may be defined in a basis independent way by the condition that
\begin{equation}
\langle\textrm{\bf{M}}^T Y,X\rangle=
\langle Y,\textrm{\bf{M}}X\rangle
\quad\textrm{for all}\quad
X\in H_m,\;Y\in H_n\;.
\end{equation}
It has matrix elements $N_{ab}=M_{ba}$, or
$N_{ik;jl}=M_{lj;ki}=(M_{jl;ik})^{\ast}$.  Thus transposition of the
map $\bf{M}$ corresponds to transposition of the real matrix $M_{ab}$
and Hermitian conjugation of the complex matrix $M_{jl;ik}$.


As defined here, the maps ${\bf M}_A$ and ${\bf M}_A^{\,T}$ act on
$X\in H_m$ and $Y\in H_n$ as follows,
\begin{equation}
\label{CJ}
{\bf M}_AX=\textrm{Tr}_1(A(X\otimes I_n))\;,\qquad
{\bf M}_A^{\,T}Y=\textrm{Tr}_2(A(I_m\otimes Y))\;,
\end{equation}
where $\textrm{Tr}_1$ and $\textrm{Tr}_2$ denote the partial traces.
The matrix $A$ may be expanded in terms of the basis vectors
$E_a\otimes F_b\in H_N$ as
\begin{equation}
A=\sum_{a,b}M_{ba}\,E_a\otimes F_b\;.
\end{equation}

\subsection{Entanglement witnesses and positive maps}

By definition, a {\em positive} map $\textrm{\bf{M}}:H_m\mapsto H_n$
transforms positive matrices into positive matrices, that is, if
$\rho\geq 0$ then $\textrm{\bf{M}}\rho\geq 0$.

In terms of the map $\textrm{\bf{M}}_A$ the biquadratic form
introduced in Equation~\eqref{eq:biqform} may be written as
\begin{equation}
f_A(\phi,\chi)
=\chi^{\dag}\,\textrm{\bf{M}}_A(\phi\phi^{\dag})\,\chi
=\phi^{\dag}\,\textrm{\bf{M}}_A^{\,T}(\chi\chi^{\dag})\,\phi\;.
\end{equation}
The condition that $f_A(\phi,\chi)\geq 0$ for all
$\phi\in\mathbb{C}^m$, $\chi\in\mathbb{C}^n$ means that
$\textrm{\bf{M}}_A(\phi\phi^{\dag})$ is a positive matrix for every
$\phi\in\mathbb{C}^m$.  Since the pure states $\phi\phi^{\dag}$ are
all the extremal points of ${\cal D}_m$, this is another way of saying
that $\textrm{\bf{M}}_A$ is a positive map.

We conclude that $A$ is an entanglement witness if and only if
$\textrm{\bf{M}}_A$ is a positive map.  An equivalent condition is
that $\textrm{\bf{M}}_A^{\,T}$ is a positive map.  This correspondence
between entanglement witnesses and positive maps places the positive
maps in a central position in the theory of quantum entanglement.

\subsection{Completely positive maps}

An obvious condition on a physical map
$\textrm{\bf{M}}:H_m\mapsto H_n\,$, transforming physical states into
physical states, is that it should be positive.  A less obvious
condition is that it should be
{\em completely positive}, so that every map of the form
$\textrm{\bf{I}}\otimes\textrm{\bf{M}}:H_k\otimes H_m\mapsto
H_k\otimes H_n$ is positive, where ${\bf{I}}:H_k\mapsto H_k$ is
the identity map.

The most general form of a completely positive map is
\begin{equation}
\textrm{\bf{M}}_AX=\sum_aV_aXV_a^{\,\dag}\;,
\end{equation}
with $n\times m$ matrices $V_a$.  This form gives the following matrix
elements of $A$,
\begin{equation}
A_{ij;kl}=(A^P)_{il;kj}=\sum_a(V_a)_{jk}(V_a)_{li}^{\,\ast}\;.
\end{equation}
We see that the map $\textrm{\bf{M}}_A$ is completely positive if and
only if $A^P$ is positive.  It is well known that $A^P$ may be
positive without $A$ being positive, because the transposition map
$\textrm{\bf{T}}X=X^T$ on $H_n$ is positive but not completely
positive.

A separable state $\rho$, as defined in Equation~\eqref{eq:sepdef},
remains positive under the application of positive maps to its local
parts, thus
\begin{equation}
({\bf{M}}\otimes {\bf{N}})\rho
=\sum_ap_a\,({\bf{M}}\sigma_a)\otimes({\bf{N}}\tau_a)
\end{equation}
is a positive matrix when ${\bf{M}}:H_m\mapsto H_k$ and
${\bf{N}}:H_n\mapsto H_l$ are positive maps.  In this way positive
maps give necessary conditions for separability.

Positive maps can also give sufficient conditions for
separability~\cite{MPRHorodecki96}.  This follows from the
correspondence described above between entanglement witnesses and
positive maps.  For
$\rho\in H_N=H_{mn}$ let us define $\sigma\in H_{m^2}$ as
\begin{equation}
\sigma=({\bf{T}}\otimes {\bf{M}}_A^{\,T})\rho\;,
\end{equation}
where $A$ is an entanglement witness and ${\bf{M}}_A$ the
corresponding positive map.  In index notation we have that
\begin{equation}
\sigma_{ka;ib}=\sum_{j,l}A_{al;bj}\,\rho_{ij;kl}\;.
\end{equation}
Clearly $\sigma$ is positive if $\rho$ is separable.  The other way
around, if $\sigma$ is positive for every entanglement witness $A$ we
have that
\begin{equation}
\textrm{Tr}(A\rho)
=\sum_{i,j,k,l}A_{kl;ij}\,\rho_{ij;kl}
=\sum_{k,a,i,b}\delta_{ka}\sigma_{ka;ib}\delta_{ib}\geq 0
\end{equation}
for every $A$, implying that $\rho$ is separable.

Note that this proof of sufficiency does not use the full condition
that $\sigma$ should be positive, it uses only the condition that
$\sigma$ should have a non-negative expectation value in the maximally
entangled pure state (Bell state) $\delta_{ik}\in\mathbb{C}^{m^2}$.
If we use the full positivity condition, one single positive map can
in principle reveal the entanglement of many different states for
which we would need many different entanglement witnesses.  The PPT
condition associated with the transposition map is a striking example,
it is equivalent to all the conditions provided by a much larger set
of entanglement witnesses.

The basic reason for the efficiency of maps in revealing entanglement
is nonlinearity: the separability condition that a product
${\bf{M}}\otimes{\bf{N}}$ of positive maps should map $\rho$ into a
positive matrix is highly nonlinear in $\rho$.  An entanglement
witness $A$, on the other hand, gives an inequality
$\textrm{Tr}(A\rho)\geq 0$ linear in $\rho$.

\subsection{Decomposable maps}

A positive map ${\bf{M}}:H_m\mapsto H_n$ is said to be
{\emph{decomposable}} if it can be written as
\begin{equation}
{\bf{M}}={\bf{M}}_1+{\bf{M}}_2{\bf{T}}_m
\end{equation}
where ${\bf{M}}_1$ and ${\bf{M}}_2$ are completely positive, and
${\bf{T}}_m$ is transposition on $H_m$~\cite{Stormer1963,Stormer1982}.


Since ${\bf M}_A{\bf T}_m={\bf M}_{A^P}$ it follows that ${\bf M}_A$
is decomposable if and only if $A=B+C$ where $B$ and $C^P$ are both
positive matrices.  This decomposition of $A$ implies for every PPT
state $\rho$ that
\begin{equation}
\textrm{Tr}(A\rho)
=\textrm{Tr}(B\rho)+\textrm{Tr}(C\rho)
=\textrm{Tr}(B\rho)+\textrm{Tr}(C^P\!\rho^P)\geq 0\;.
\end{equation}
In particular, $A$ is an entanglement witness, and we call it a
decomposable witness.  It is not very useful as a witness, however,
since it can not reveal the entanglement of an entangled PPT state.

It was shown by Woronowicz that all positive maps
${\bf{M}}:H_m\mapsto H_n$ with $N=mn\leq 6$ are
decomposable~\cite{Woronowicz1976}.  In these dimensions entangled PPT
states therefore do not exist, and the Peres criterion is both
necessary and sufficient for separability.  By virtue of the Peres
criterion, the development of practically useful separability criteria
in dimensions $N\geq 8$ is a problem closely related to the
understanding of the nondecomposable positive maps.

\subsection{Unital and trace preserving maps}
\label{sec:Unitaltrprmaps}

A map ${\bf M}:H_m\mapsto H_n$ is {\em unital} if it maps the identity
$I_m\in H_m$ to the identity $I_n\in H_n$, that is, if
\begin{equation}
\sum_{i,k}M_{jl;ik}\,\delta_{ik}=
\sum_iM_{jl;ii}=\delta_{jl}\;.
\end{equation}
It is {\em trace preserving} if
$\textrm{Tr}(\textrm{\bf{M}}X)=\textrm{Tr}\,X$ for all $X\in H_m$,
that is, if
\begin{equation}
\sum_{j,i,k}M_{jj;ik}X_{ik}=\sum_iX_{ii}\;,
\end{equation}
which means that
\begin{equation}
\sum_jM_{jj;ik}=\delta_{ik}\;.
\end{equation}
Thus the map $\textrm{\bf{M}}$ is trace preserving if and only if
$\textrm{\bf{M}}^T$ is unital.
A physical map should conserve probability, which means that it should
be trace preserving.

In terms of the corresponding matrix
$A\in H_N$ the condition for $\textrm{\bf{M}}_A$ to be unital is that
\begin{equation}
\label{eq:Munital}
\textrm{\bf{M}}_AI_m=\textrm{Tr}_1A=I_n\;,
\end{equation}
and the condition for $\textrm{\bf{M}}_A$ to be trace preserving is
that
\begin{equation}
\label{eq:Mtracepres}
\textrm{\bf{M}}_A^{\,T}I_n=\textrm{Tr}_2A=I_m\;.
\end{equation}
%
Equation~\eqref{eq:Munital} implies that
$\textrm{Tr}\,A=\textrm{Tr}\,I_n=n$, whereas
Equation~\eqref{eq:Mtracepres} implies that
$\textrm{Tr}\,A=\textrm{Tr}\,I_m=m$.  Thus the map may be both unital
and trace preserving only if $m=n$.  If $m\neq n$ the proper condition
is rather that
\begin{equation}
\label{eq:unitaltrprM}
\textrm{\bf{M}}_AE_0=F_0\;,\qquad
\textrm{\bf{M}}_A^{\,T}F_0=E_0\;,
\end{equation}
where $E_0=I_m/\sqrt{m}$ and $F_0=I_n/\sqrt{n}\,$.  Since we regard
normalization constants as unimportant we will abuse the language and
call the map unital and trace preserving if
Equation~\eqref{eq:unitaltrprM} holds, even when $m\neq n$.

In terms of the matrix elements
$M_{ba}=\langle F_b,\textrm{\bf{M}}_AE_a\rangle$ the condition for
unitality is that $M_{b0}=\delta_{b0}$, and the condition for trace
preservation is that $M_{0a}=\delta_{0a}$.  This means that a unital
and trace preserving map $\textrm{\bf{M}}_A$ is generated by an
entanglement witness of the form
\begin{equation}
\label{eq:unitaltrprA}
A=\frac{I_N}{\sqrt{N}}
+\sum_{a=1}^{m^2-1}\sum_{b=1}^{n^2-1}M_{ba}\,E_a\otimes F_b\;.
\end{equation}

\subsection{Zeros of entanglement witnesses}


Since the conditions defining $A\in H_N$ as an entanglement witness
are the infinite set of inequalities
\begin{equation}
f_A(\phi,\chi)\geq 0
\qquad
\textrm{for all}
\qquad
\phi\in\mathbb{C}^m\;,\;
\chi\in\mathbb{C}^n\;,
\end{equation}
it is clear that $A$ is a boundary point of ${\cal S}^{\circ}$ if and
only if at least one of these inequalities is an equality.
We call the pure product state $\psi_0\psi_0^{\dag}$ with
$\psi_0=\phi_0\otimes\chi_0$ and
$\phi_0^{\dag}\phi_0=\chi_0^{\dag}\chi_0=1$ a zero of $A$ if
\begin{equation}
\label{eq:wzero}
f_A(\phi_0,\chi_0)=0\;.
\end{equation}
Since a zero is a minimum of the non-negative function $f_A$, it
follows that the first derivative in every direction at the zero must
also vanish.  These conditions amount to a set of equalities
\begin{equation}
(\phi_0\otimes\chi_0)^{\dag}A(\phi\otimes\chi_0)=
(\phi_0\otimes\chi_0)^{\dag}A(\phi_0\otimes\chi)=0
\qquad
\textrm{for all}
\qquad
\phi\in\mathbb{C}^m\;,\;
\chi\in\mathbb{C}^n\;.
\end{equation}
These equations may be regarded as linear constraints on the matrix
$A$.  It should be remembered that each equation is complex and is
equivalent to two real equations, except that
Equation~\eqref{eq:wzero} is real.  A careful counting shows that one
zero implies $2(m+n)-3$ real valued constraints on $A$.

If the second derivative in every direction is strictly positive, then
the zero $(\phi_0,\chi_0)$ is a quadratic minimum, and there are no
more constraints on $A$ from this zero.  If the second derivative in
some direction vanishes, then the third derivative in this direction
must also vanish, and the zero is a quartic minimum.  This imposes
further constraints on $A$, which we do not detail here.

The distinction between quadratic and quartic zeros is important.
Obviously, the quadratic zeros are isolated points.  A quartic zero
may be an isolated point, but it may also belong to a continuous set
of zeros.

An entanglement witness $A$ is extremal if and only if it has so many
zeros that all the constraints from all the zeros together determine
$A$ uniquely up to a proportionality constant.

\subsection{Extremal positive maps}

An extremal entanglement witness $A$ is understood in terms of its
zeros.  It corresponds to an extremal positive map
$\textrm{\bf{M}}_A$, and we have to understand what the zeros of the
witness imply for the corresponding map.

Let $(\phi_0,\chi_0)$ be an isolated zero of $A$, and define
$Y=\textrm{\bf{M}}_A(\phi_0\phi_0^{\dag})\,$,
$X=\textrm{\bf{M}}_A^{\,T}(\chi_0\chi_0^{\dag})\,$.  We use the
identities
\begin{equation}
\chi_0^{\dag}Y\chi_0
=\phi_0^{\dag}X\phi_0
=(\phi_0\otimes\chi_0)^{\dag}A\,(\phi_0\otimes\chi_0)=0\;.
\end{equation}
Since $Y$ and $X$ are both positive matrices, it follows that
\begin{equation}
Y\chi_0=0\;,\qquad
X\phi_0=0\;.
\end{equation}
We also know that $Y$ and $X$ have no other zero vectors, otherwise
the zero vectors would span subspaces of dimension two or higher, and
the zero $(\phi_0,\chi_0)$ would not be isolated.  Hence $Y$ has rank
$n-1$ and $X$ has rank $m-1$.

To summarize, when $A$ is an extremal entanglement witness the
corresponding maps $\textrm{\bf{M}}_A$ and $\textrm{\bf{M}}_A^{\,T}$
are extremal positive maps.  An isolated zero $(\phi_0,\chi_0)$ of $A$
defines a rank one state $\phi_0\phi_0^{\dag}\in{\cal D}_m$ mapped by
$\textrm{\bf{M}}_A$ to a rank $n-1$ state in ${\cal D}_n$, and a rank
one state $\chi_0\chi_0^{\dag}\in{\cal D}_n$ mapped by
$\textrm{\bf{M}}_A^{\,T}$ to a rank $m-1$ state in ${\cal D}_m$.

Thus, the zero of $A$ defines a point on the boundary of ${\cal D}_m$
which is mapped to a point on the boundary of ${\cal D}_n$, and a
point on the boundary of ${\cal D}_n$ which is mapped by the
transposed map to a point on the boundary of ${\cal D}_m$.  We
understand that the map $\textrm{\bf{M}}_A$ is extremal precisely
because the image $\textrm{\bf{M}}_A{\cal D}_m$ inside ${\cal D}_n$
touches the boundary of ${\cal D}_n$ in as many points as possible.

\subsection{Rank one preserving maps}

The identity map ${\bf I}:X\mapsto X$ and the transposition map
${\bf T}:X\mapsto X^T=X^{\ast}$ are two examples of extremal positive
maps.  They have the special property that they are rank one
preservers, mapping pure states to pure states.  They are extremal
precisely because they map the boundary of $\mathcal{D}$ onto itself.
They both preserve volume.  In fact, the eigenvalues of ${\bf T}$ are
$\pm 1$, because ${\bf T}$ is its own inverse, ${\bf T}^2={\bf I}$.

More generally, let $U$ be a unitary matrix and define
${\bf U}:X\mapsto UXU^{\dagger}$.  Then ${\bf U}$ and ${\bf UT}$ are
both rank one preservers, and they are extremal because they map the
boundary of $\mathcal{D}$ onto itself.  They also preserve volumes.

These maps and the corresponding entanglement witnesses are
fundamentally different from the generic extremal witnesses we find
numerically, which have only quadratic zeros, so that the
corresponding maps map only a finite number of points from the
boundary of $\mathcal{D}$ to the boundary, and map all other boundary
points to the interior.  An extremal map of this generic kind is
contractive, it reduces volumes, since it maps $\mathcal{D}$ to a
subset of itself.

\section{Transforming positive maps to unital and trace preserving form}
\label{sec:Transforming}

We argue now that every positive map ${\bf{M}}={\bf{M}}_A$ where $A$
is an entanglement witness may be transformed into a unital and trace
preserving form through a product transformation of the form
\cite{Stormer2013}
\begin{equation}
  \label{CJ-Jan}
  A\mapsto \widetilde{A}=(U\otimes V)A(U\otimes V)^{\dagger}
\end{equation}
where $U\in\mbox{GL}(m,\mathbb{C})$ and $V\in\mbox{GL}(n,\mathbb{C})$.
Furthermore, we present here an efficient iteration procedure for
doing the transformation numerically.

A product transformation of this kind preserves all the essential
characteristics of $A$.  For instance, if $A$ is extremal in
$\mathcal{S}^{\circ}$ and non-decomposable, then so is
$\widetilde{A}$, and a zero $\phi\otimes \chi$ of $A$ corresponds to a
zero $\widetilde{\phi}\otimes\widetilde{\chi}$ of $\widetilde{A}$,
where
\begin{equation}
\widetilde{\phi}={(U^{\dagger})}^{-1}\phi\;,
\qquad
\widetilde{\chi}={(V^{\dagger})}^{-1}\chi\;.
\end{equation}

As we saw in Section~\ref{sec:Unitaltrprmaps} the result to be proved
is that every entanglement witness $A$ may be transformed to a form as
given in Equation~\eqref{eq:unitaltrprA}.  In~\cite{Leinaas2006} it
was proved that every strictly positive density matrix may be transformed to
such a form.  But the theorem is actually valid more generally than it
is stated there, since the proof is valid for every entanglement
witness having no zeros, in other words, every witness lying in the
interior of $\mathcal{S}^{\circ}$.  Here we want to apply this type of
transformation to extremal witnesses, which lie on the boundary
$\partial\mathcal{S}^{\circ}$ and have a maximal number of zeros.
What could in principle go wrong in the limit when the boundary is
approached from the inside of $\mathcal{S}^{\circ}$ is that the
transformation could become singular, but we find in practice that no
such problems arise.

In terms of the transformed map
$\widetilde{{\bf{M}}}={\bf{M}}_{\widetilde{A}}$ the conditions to be
fulfilled are that $\widetilde{{\bf{M}}}I_m=I_n$ and
$\widetilde{{\bf{M}}}^TI_n=I_m$.  We will assume here that $m=n$,
otherwise the proper conditions would be that
$\widetilde{{\bf{M}}}(I_m/\sqrt{m})=I_n/\sqrt{n}$ and
$\widetilde{{\bf{M}}}^T(I_n/\sqrt{n})=I_m/\sqrt{m}$.

In index notation Equation~\eqref{CJ-Jan} reads as follows,
\begin{equation}
\widetilde{A}_{ij;kl}
=\sum_{a,b,c,d} U_{ia}V_{jb}\,A_{ab;cd}\,U^{*}_{kc}V^{*}_{ld}\;.
\end{equation}
According to Equation~\eqref{CJ} the transformation
$Y=\widetilde{\textrm{\bf{M}}}X$ then reads as follows,
\begin{equation}
Y_{jl}=\sum_{i,k}\widetilde{A}_{ij;kl}\,X_{ki}
=\sum_{i,k,a,b,c,d}V_{jb}(A_{ab;cd}(U^{*}_{kc}\,X_{ki}\,U_{ia}))V^*_{ld}\;,
\end{equation}
or in the indexfree notation,
\begin{equation}
Y=\widetilde{\textrm{\bf{M}}}X
=V(\textrm{\bf{M}}(U^{\dagger}XU))V^{\dagger}\;.
\end{equation}
Similarly, the transformation $X=\widetilde{\textrm{\bf{M}}}^TY$ reads
\begin{equation}
X_{ik}=\sum_{j,l}\widetilde{A}_{ij;kl}Y_{lj}
=\sum_{j,l,a,b,c,d}U_{ia}(A_{ab;cd}(V^*_{ld}Y_{lj}V_{jb}))U^{*}_{kc}
\;,
\end{equation}
or
\begin{equation}
X=\widetilde{\textrm{\bf{M}}}^TY
=U(\textrm{\bf{M}}^T(V^{\dagger}YV))U^{\dagger}\;.
\end{equation}
The conditions for $\widetilde{\textrm{\bf{M}}}$ to be unital and
trace preserving are that
\begin{align}
\widetilde{\textrm{\bf{M}}}I_m
&=
V(\textrm{\bf{M}}(U^{\dagger}U))V^{\dagger} = I_n\;,
\nonumber
\\
\widetilde{\textrm{\bf{M}}}^T\!I_n
&=
U(\textrm{\bf{M}}^T\!(V^{\dagger}V))U^{\dagger}=I_m\;.
\end{align}
Thus the problem to be solved is to find $U$ and $V$ such that
\begin{equation}
\textrm{\bf{M}}(U^{\dagger}U) = (V^{\dagger}V)^{-1}\;,
\qquad
\textrm{\bf{M}}^T\!(V^{\dagger}V) = (U^{\dagger}U)^{-1}\;.
\end{equation}

\subsection*{Solution by iteration}

This problem can be solved in two steps.  First we find positive
Hermitian matrices $X=U^{\dagger}U$ and $Y=V^{\dagger}V$ solving the
equations
\begin{equation}
\label{eq:XY}
\textrm{\bf{M}}X = Y^{-1}\;,
\qquad
\textrm{\bf{M}}^TY = X^{-1}\;.
\end{equation}
Then we solve the equations
\begin{equation}
U^{\dagger}U=X\;,\qquad
V^{\dagger}V=Y
\end{equation}
for $U$ and $V$.  The general solutions are
\begin{equation}
U=U_1U_2\;,\qquad
V=V_1V_2\;,
\end{equation}
where $U_1$ and $V_1$ are arbitrary unitary matrices, and
$U_2=\sqrt{X}$, $V_2=\sqrt{Y}$ are the uniquely defined positive
Hermitian square roots, which we compute for example by diagonalizing
$X$ and $Y$ and taking the square roots of the eigenvalues.

Equation~\eqref{eq:XY} makes sense because the matrices
$X=U^{\dagger}U$ and $Y=V^{\dagger}V$ are strictly positive as long as
$U$ and $V$ are nonsingular, and the maps ${\bf M}$ and ${\bf M}^T$,
as well as the inversions $X\mapsto X^{-1}$ and $Y\mapsto Y^{-1}$,
transform strictly positive matrices into strictly positive matrices.

The method suggesting itself for solving Equation~\eqref{eq:XY} is
simply to iterate the equations.  Given an approximate solution $X_k$
for $X$ we try to compute a better approximation $X_{k+1}$ by a series
of four transformations,
\begin{equation}
X_k\mapsto
S_k={\textrm{\bf{M}}}X_k\mapsto
Y_k=S_k^{-1}\mapsto
T_k=\textrm{\bf{M}}^TY_k\mapsto
X_{k+1}=T_k^{-1}\;.
\end{equation}
We start the iterations for example with $X_0=I$.

A sufficient condition for the convergence of $X_k$ to a unique limit
$X$ is that each transformation $X_k\mapsto X_{k+1}$ is contractive
(except in the direction along $X$).  A small perturbation
$\Delta X_k$ of $X_k$ transforms linearly as
\begin{equation}
\Delta X_{k+1}={\bf D}(\Delta X_k)\;,
\end{equation}
where the linear map ${\bf D}$ is the derivative of the nonlinear
transformation $X_k\mapsto X_{k+1}$.  The transformation is
contractive if all eigenvalues of ${\bf D}$ (except the special
eigenvalue which must be close to 1) are smaller than one in
absolute value.  Now ${\bf D}$ is a composition of four linear maps,
\begin{equation}
{\bf D}={\bf D}_4{\bf M}^T{\bf D}_2{\bf M}\;,
\end{equation}
where ${\bf D}_2$ and ${\bf D}_4$ are linearizations of the
matrix inversions,
\begin{align}
{\bf D}_2(\Delta S_k)&=-S_k^{-1}(\Delta S_k)S_k^{-1}\;,
\nonumber\\
{\bf D}_4(\Delta T_k)&=-T_k^{-1}(\Delta T_k)T_k^{-1}\;.
\end{align}

The examples of extremal positive maps ${\bf M}$ and ${\bf M}^T$ that
we have studied numerically are strongly contractive, and we find in
practice that this is enough to ensure that ${\bf D}$ is contractive,
with eigenvalues typically no larger than about $0.5$, even though
${\bf D}_2$ and ${\bf D}_4$ are not contractive.

We have used this iteration scheme on a large number of numerically
produced extremal entanglement witnesses \cite{Hansen2014}, and also
on many non-extremal witnesses constructed as convex combinations of
the extremal ones.  Numerically our attempts, which are in the
thousands, always converge, and it also appears that for a given
witness $A$ the solution $X$ is unique, independent of the initial
guess $X_0$.

\section{Visualization of positive maps}

We now specialize to the case $m=n=3, N=mn=9$.  The set
$\mathcal{D}=\mathcal{D}_3$ of normalized density matrices has
dimension $3^2-1=8$.  When $m=n$ a linear positive map from $H_m$ to
$H_n$ may be transformed to a form where it is unital and trace
preserving.

Given a unital and trace preserving positive map
${\bf{M}}:H_3\mapsto H_3$.  It maps $\mathcal{D}$ into $\mathcal{D}$,
and the maximally mixed state $I/3$ to itself.  We plot two
dimensional planar sections in order to illustrate how the image
${\bf{M}}\mathcal{D}$ lies inside $\mathcal{D}$.

We present such plots here for two different examples of extremal
positive maps.  The first example corresponds to a randomly chosen
generic extremal witness found in a numerical search.  Such witnesses
have only quadratic zeros.  The second example is the Choi--Lam map,
which is qualitatively different since the corresponding witness has
only quartic zeros.

\subsection{Two dimensional sections through the set of density
  matrices}

In most of our plots we use three density matrices
$\rho_0,\rho_1,\rho_2\in\mathcal{D}$ to define a plane
$\mathcal{Z}\subset H_3$.  We use $\rho_0$ as origin in the plane, and
define coordinate axes
\begin{equation}
\label{eq:ABaxes}
B=a(\rho_1-\rho_0)\;,
\qquad
C=b(\rho_2-\rho_0)+c(\rho_1-\rho_0)\;,
\end{equation}
with real constants $a,b,c$ chosen in some way to be specified later
on.  We will always choose $a>0$ and $b>0$.  A matrix
$X\in\mathcal{Z}$ is then specified by a coordinate pair $(x,y)$ as
\begin{equation}
\label{XxAyB}
X=\rho_0+xB+yC\;.
\end{equation}
Note that $\textrm{Tr}\,B=\textrm{Tr}\,C=0$ because
$\textrm{Tr}\,\rho_i=1$ for $i=0,1,2$, hence
$\textrm{Tr}\,X=\textrm{Tr}\,\rho_0=1$.

\subsection*{Six types of sections}

\begin{figure}[H]
\begin{center}
  \includegraphics[width=0.7\textwidth]{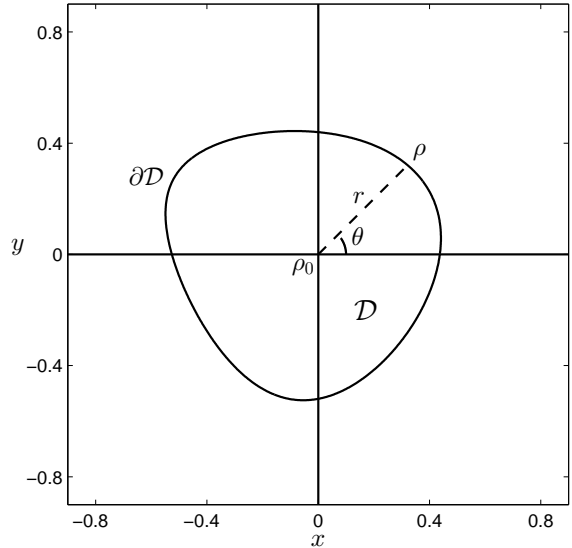}
  \caption{\label{fig:1} A two dimensional section through
    $\mathcal{D}$, the set of density matrices. The curve represents
    the boundary $\partial\mathcal{D}$ which consists of the states of
    less than full rank. For a given angle $\theta$ the distance $r$
    is calculated numerically. The state $\rho_0=I/3$ is the origin,
    and the state $\rho$ on the boundary has rank two.}
\end{center}
\end{figure}

Figure~\ref{fig:1} is an example of a section through the set of
density matrices, where we have chosen the maximally mixed state as
the origin, that is, $\rho_0=I/3$.  The density matrices $\rho_1$ and
$\rho_2$ helping to define the section are here chosen at random.  By
Equation~\eqref{eq:ABaxes} the $x$ axis goes in the direction from
$\rho_0$ to $\rho_1$.  For this plot we have chosen the constants
$a,b,c$ in Equation~\eqref{eq:ABaxes} such that $B$ and $C$ are
orthogonal unit vectors in $H_3$, that is,
\begin{equation}
\textrm{Tr}(B^2)=\textrm{Tr}(C^2)=1\;,\qquad
\textrm{Tr}(BC)=0\;.
\end{equation}
This means that distances in our plot represent faithfully distances
in $H_3$ as defined by the Hilbert--Schmidt metric.  To locate
numerically the boundary $\partial\mathcal{D}$ we write
$x=r\cos\theta$, $y=r\sin\theta$, and with $\theta$ fixed we determine
the largest value of $r$ such that the matrix $X=\rho_0+xB+yC$ has no
negative eigenvalues.  The boundary is here a smooth curve consisting
entirely of rank two states.

\begin{figure}[H]
  \includegraphics[width=0.36\textwidth]{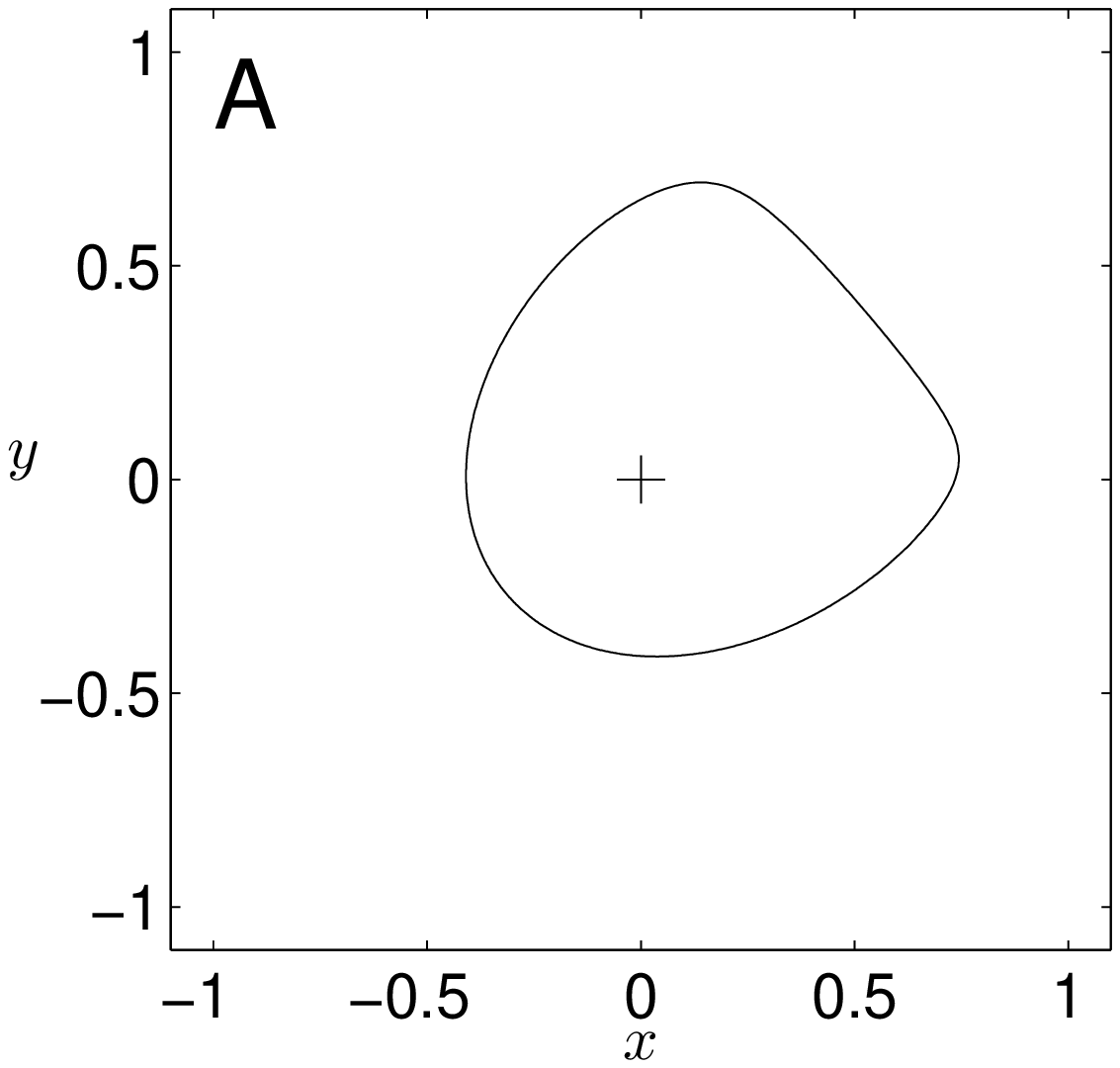}
  \hspace*{-0.06\textwidth}
  \includegraphics[width=0.36\textwidth]{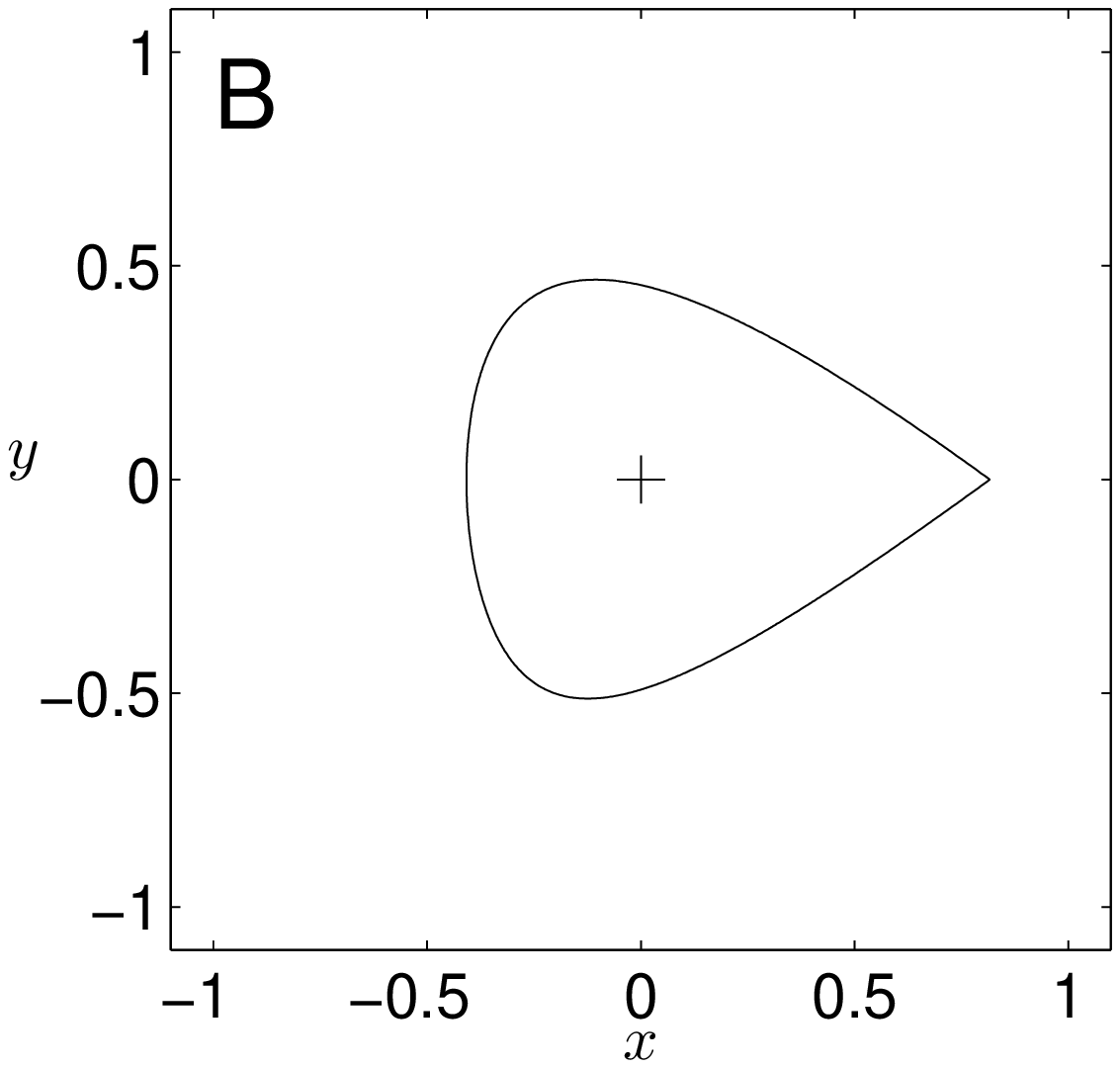}
  \hspace*{-0.06\textwidth}
  \includegraphics[width=0.36\textwidth]{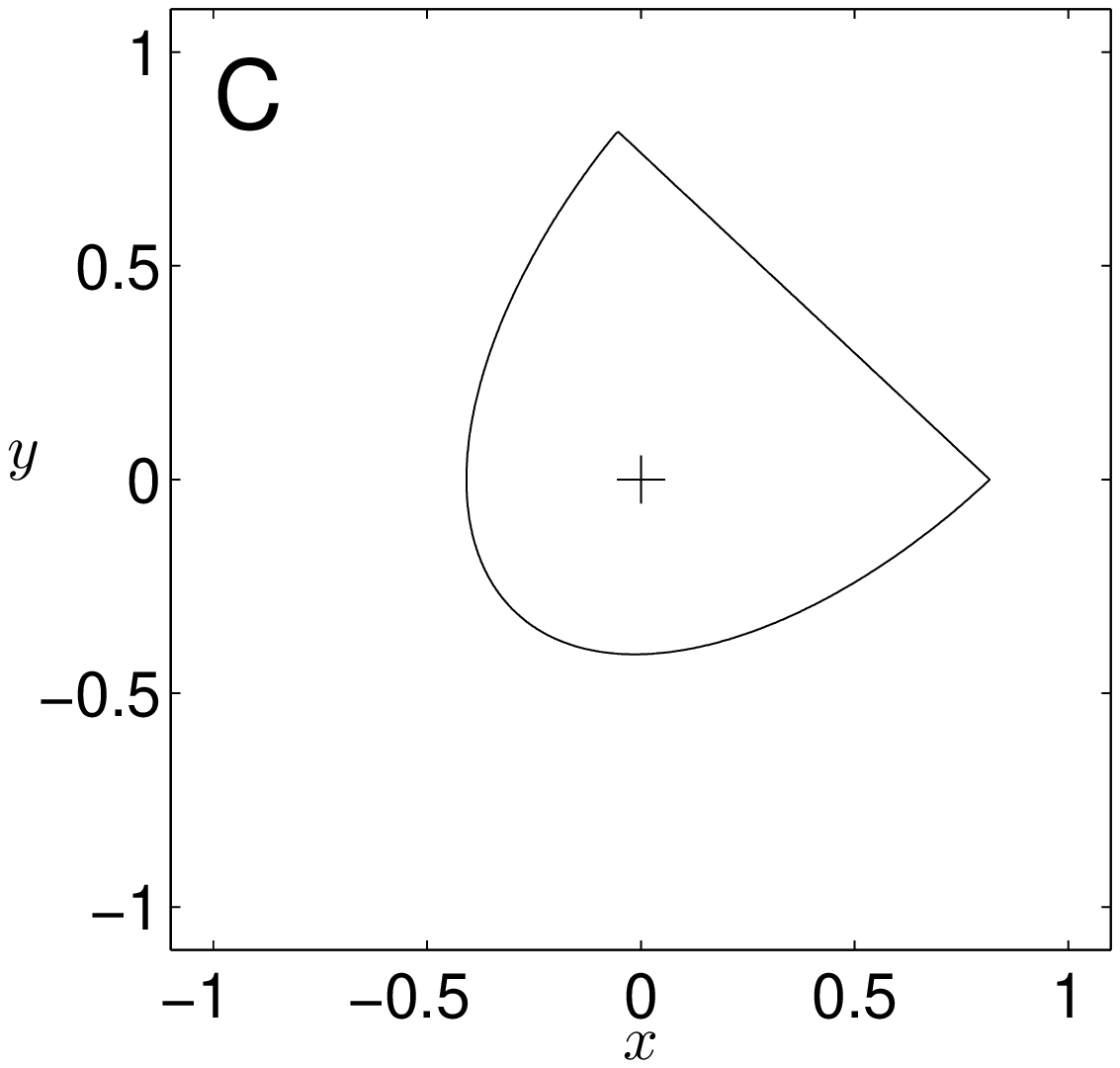}

  \includegraphics[width=0.36\textwidth]{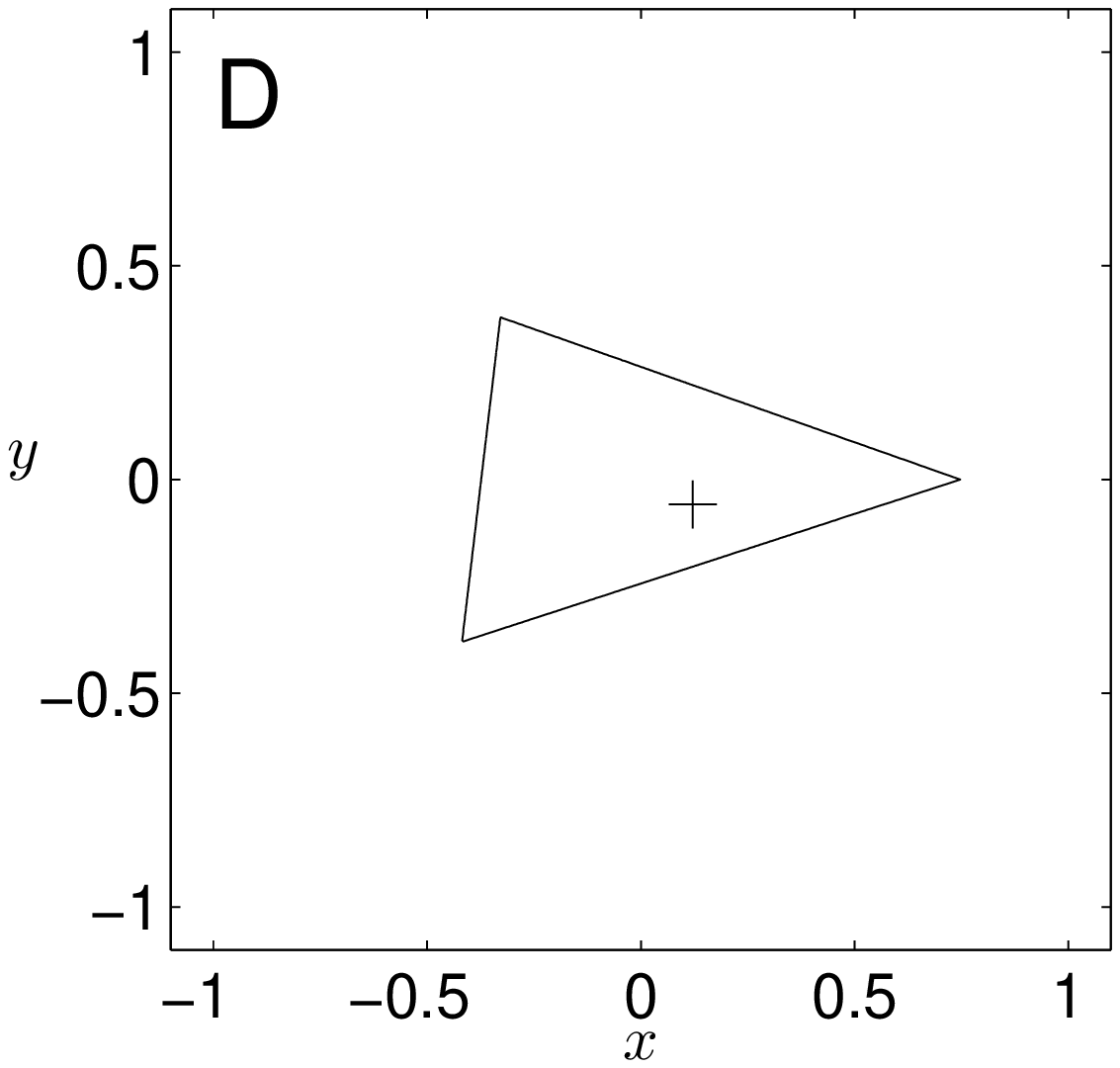}
  \hspace*{-0.06\textwidth}
  \includegraphics[width=0.36\textwidth]{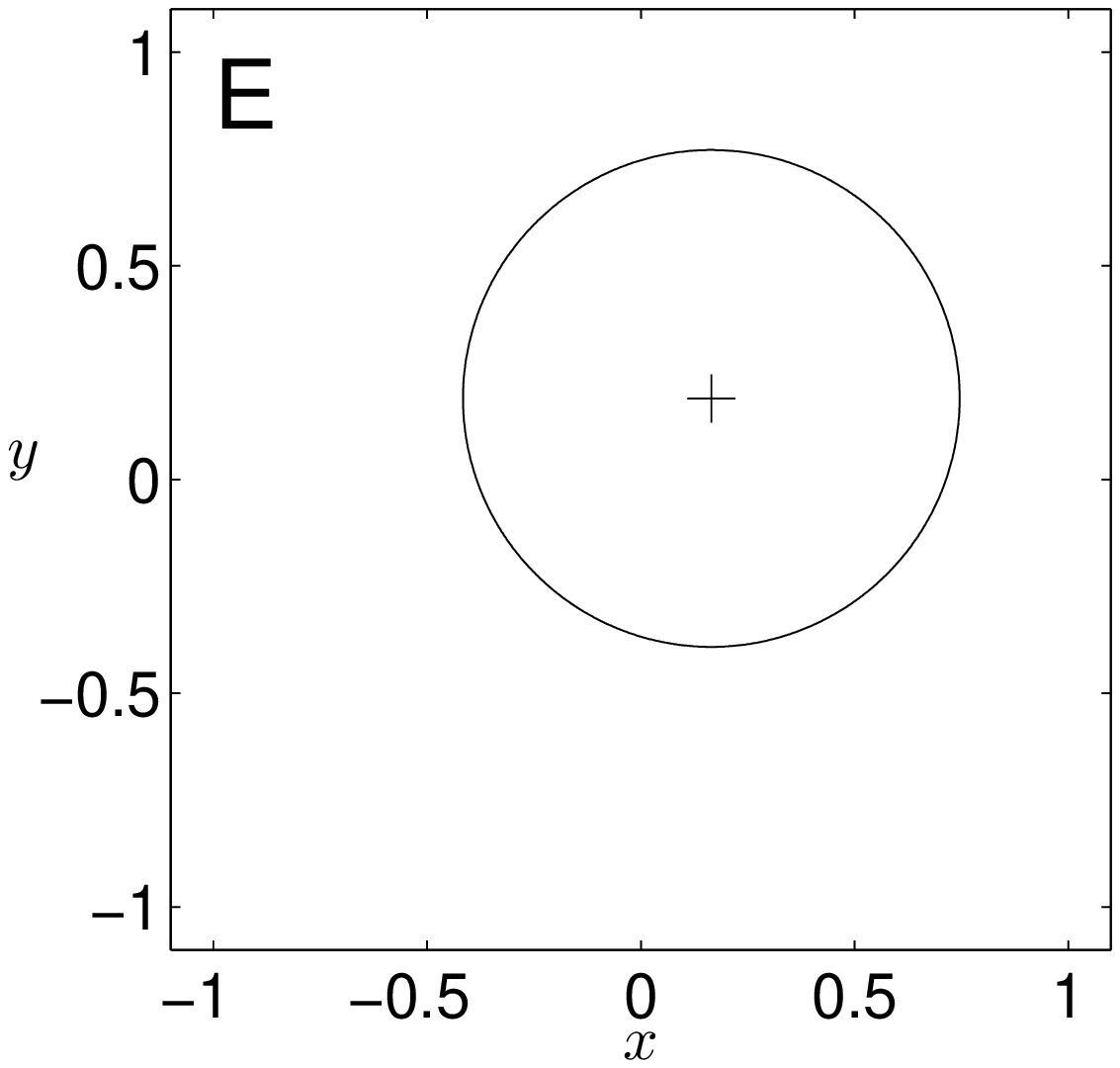}
  \hspace*{-0.06\textwidth}
  \includegraphics[width=0.36\textwidth]{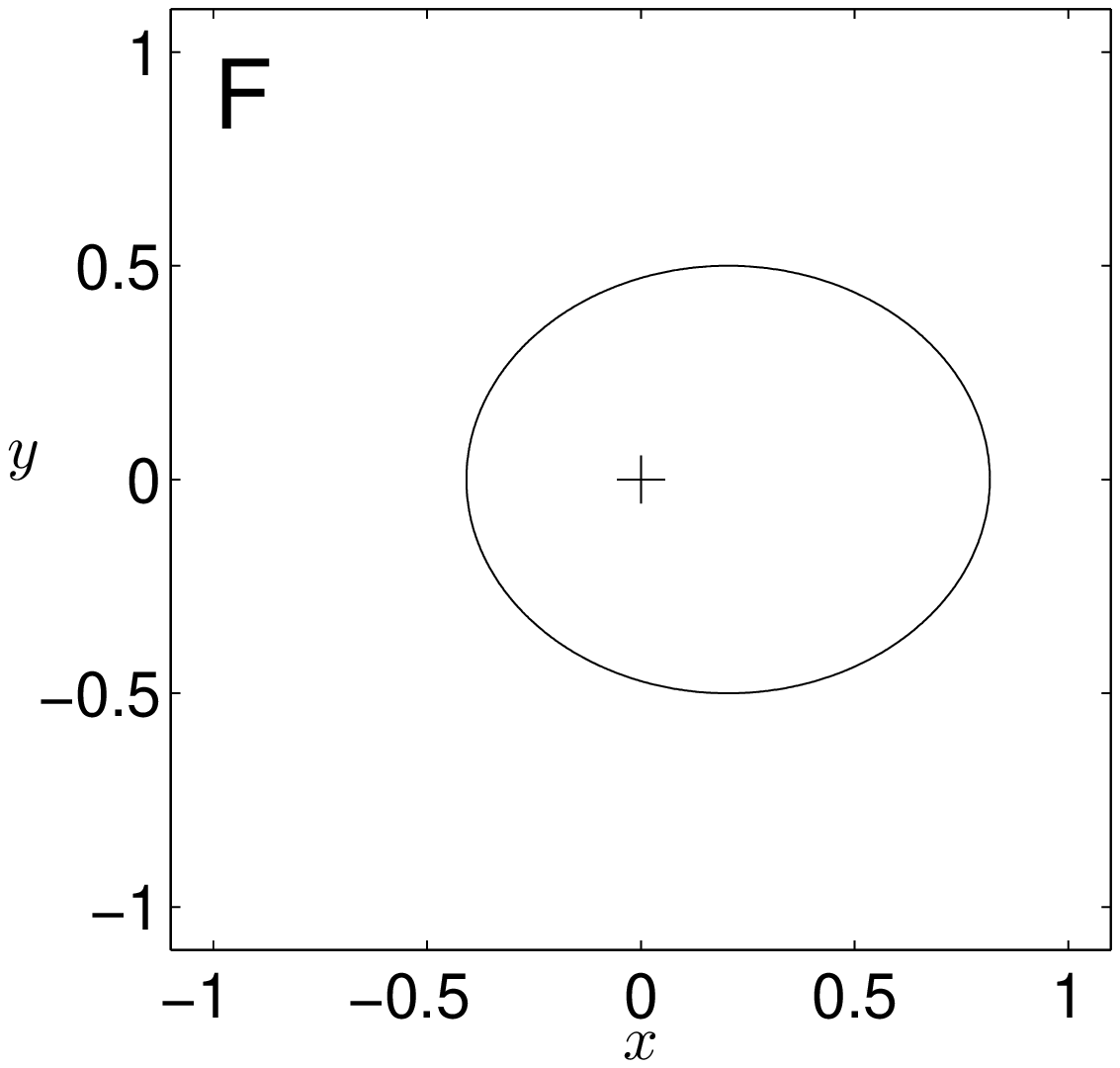}


  \caption{\label{fig:illustrasjon}Six types of two dimensional
    sections through $\mathcal{D}$, as explained in the text.}

\end{figure}

Figure~\ref{fig:illustrasjon} shows six types of sections of
$\mathcal{D}$.  In the sections~A, B, C, and F the origin $\rho_0$ is
the maximally mixed state $I/3$, it is marked by a ``$+$''.  In the
sections D and E the ``$+$'' is the orthogonal projection of the
maximally mixed state.  All sections, except section~E, cut through
the interior of $\mathcal{D}$, so that the interior points in the
sections are states of the full rank three.

Section~A is of the same type as Figure~\ref{fig:1}, here $\rho_1$ and
$\rho_2$ are chosen as random states of rank two, and the boundary is
a smooth curve of rank two states.

In section~B we choose $\rho_1$ to be a pure state, while $\rho_2$ is
a random state of full rank.  The pure state is seen in the plot as
the single point on the boundary curve where the tangent direction is
discontinuous.  All other boundary points are states of rank two.

In section~C both $\rho_1$ and $\rho_2$ are pure states, they are
joined in the plot by a straight line of rank two states.  The curved
part of the boundary also consists of rank two states.

Section~D is a simplex where the three corners are pure states
$\rho_k=\phi_k\phi_k^{\dagger}$ for $k=1,2,3$ defined by linearly
independent vectors $\phi_k\in\mathbb{C}^3$.  The straight lines
joining the pure states consist of rank two states.  The origin is
chosen as an even mix of the three pure states.

Section~E is a two dimensional section through a three dimensional
Bloch sphere contained in the boundary $\partial\mathcal{D}$.  Like
section~D it is defined by three pure states, but in this case the
three vectors in $\mathbb{C}^3$ are linearly dependent.  In this
section the interior points have rank two, and the boundary curve is a
circle of pure states.

In section~F we choose again a pure state
$\rho_1=\phi_1\phi_1^{\dagger}$, but a section in which the boundary
curve is smooth at this point.  The three matrices defining the
section are $\rho_0=I/3$, $\rho_1$, and $\rho_2=\rho_1+D$, where the
matrix $D$ is chosen in such a way that $\rho_1+\epsilon D$ for
$\epsilon$ real is a pure state to first order in $\epsilon$, that is,
\begin{equation}
(\phi_1+\epsilon\xi)(\phi_1+\epsilon\xi)^{\dagger}
=\rho_1+\epsilon D+\mathcal{O}(\epsilon^2)
\end{equation}
for some $\xi\in\mathbb{C}^3$.
We see that the proper choice is
\begin{equation}
D=\phi_1\xi^{\dagger}+\xi\phi_1^{\dagger}\;.
\end{equation}
It is of no importance that $\rho_2\not\in\mathcal{D}$.  We choose
$\xi$ orthogonal to $\phi_1$ in order to have $\textrm{Tr}\,D=0$.  The
boundary points in this section, apart from $\rho_1$, are rank two
states.

\subsection*{Visualizing maps}

The positive map $\textrm{\bf{M}}$ maps the plane $\mathcal{Z}$
defined by the states $\rho_i\in\mathcal{D}$ with $i=0,1,2$ into the
plane $\widetilde{\mathcal{Z}}={\bf{M}}\mathcal{Z}$ defined by the
states $\widetilde{\rho}_i=\textrm{\bf{M}}\rho_i\in\mathcal{D}$, in
such a way that $X$ as given in Equation~\eqref{XxAyB} is mapped into
\begin{equation}
\label{MXxMAyMB}
\widetilde{X}={\bf{M}}X
=\widetilde{\rho}_0+x\widetilde{B}+y\widetilde{C}\;,
\end{equation}
with
\begin{equation}
\widetilde{B}
=a(\widetilde{\rho}_1-\widetilde{\rho}_0)\;,
\qquad
\widetilde{C}
=b(\widetilde{\rho}_2-\widetilde{\rho}_0)
+c(\widetilde{\rho}_1-\widetilde{\rho}_0)\;.
\end{equation}
It is an important point that
$\textrm{Tr}(\widetilde{B})=\textrm{Tr}(\widetilde{C})=0$.  This
follows because we require the map ${\bf M}$ to be trace preserving.

The constants $a,b,c$ are the same here as in
Equation~\eqref{eq:ABaxes}.  We choose them now to have such values
that $\widetilde{B}$ and $\widetilde{C}$ are orthogonal unit vectors,
that is,
\begin{equation}
\textrm{Tr}(\widetilde{B}^2)=\textrm{Tr}(\widetilde{C}^2)=1\;,\qquad
\textrm{Tr}(\widetilde{B}\widetilde{C})=0\;.
\end{equation}
The motivation for this choice is that we want our plots of the
coordinates $(x,y)$ to represent faithfully the distances in the image
plane $\widetilde{\mathcal{Z}}$ of the map ${\bf M}$.  These values
for $a,b,c$ will obviously not, as a rule, be the same values that
make $B$ and $C$ orthogonal unit vectors.  Thus our $(x,y)$ plot will
be a distorted representation of the plane $\mathcal{Z}$.  The
distortion is of course due to the map ${\bf M}$ between the two
planes.

To summarize, in one and the same plot the coordinate pair $(x,y)$
represents both the matrix $X\in\mathcal{Z}$, given by
Equation~\eqref{XxAyB}, and the matrix
$\widetilde{X}={\bf{M}}X\in\widetilde{\mathcal{Z}}$, given by
Equation~\eqref{MXxMAyMB}.  Note that our definition of the axes $B,C$
and $\widetilde{B},\widetilde{C}$ is such that $\widetilde{\rho}_1$
always has coordinates $y=0$ and $x>0$ in the plots, whereas
$\widetilde{\rho}_2$ has $y>0$ but in general $x\neq 0$.

We plot the boundary of $\mathcal{D}$ in $\widetilde{\mathcal{Z}}$ by
a full drawn curve, and the boundary of $\mathcal{D}$ in $\mathcal{Z}$
by a dashed curve.  We calculate these boundaries by the method
already described.  That is, we write $x=r\cos\theta$,
$y=r\sin\theta$.  For a fixed value of $\theta$ we determine $r_1$ as
the largest value of $r$ such that $X$ has no negative eigenvalue,
this defines a point $(x_1,y_1)$ on the boundary of $\mathcal{D}$ in
$\mathcal{Z}$, or equivalently, on the boundary of
${\bf{M}}\mathcal{D}$ in $\widetilde{\mathcal{Z}}$.  And we determine
$r_2$ as the largest value of $r$ such that $\widetilde{X}$ has no
negative eigenvalue, this defines a point $(x_2,y_2)$ on the boundary
of $\mathcal{D}$ in $\widetilde{\mathcal{Z}}$.

\subsection{One example: A generic extremal non-decomposable positive map}

In dimensions $m=n=3$, $N=mn=9$, we have constructed numerically
extremal entanglement witnesses by random searches~\cite{Hansen2014}.
By definition, extremal witnesses found in random searches are
generic.

The generic extremal witnesses we find have only quadratic zeros.
Recall that an extremal witness is uniquely determined by its zeros.
Since the number of constraints on the witness from one quadratic zero
is $2(m+n)-3=9$, and the number of independent constraints must be
$N^2-1=80$, at least nine zeros are necessary to determine the witness
uniquely.  Most of our generic extremal witnesses have exactly nine
quadratic zeros, giving a total of 81 constraints, of which 80 are
then independent.

The plots presented here show the action of one unital and trace
preserving map ${\bf{M}}={\bf{M}}_A$, where $A$ is the transformed
version, as described in Section~\ref{sec:Transforming}, of a randomly
selected generic extremal entanglement witness from the sample
described in~\cite{Hansen2014}.  The zeros of the transformed witness
$A$ are $\psi_i=\phi_i\otimes\chi_i$ with $i=1,2,\ldots,9$.

The ``$+$'' in each plot represents the orthogonal projection of the
maximally mixed state $I/3$ on the image plane
$\widetilde{\mathcal{Z}}={\bf{M}}\mathcal{Z}$.  Since we use unital
maps, if we choose $\rho_0=I/3$ then
$\widetilde{\rho}_0={\bf{M}}\rho_0=\rho_0$ and the ``$+$'' is at the
origin of the plot.  The solid curve in each plot represents the
boundary $\partial\mathcal{D}$ in the image plane
$\widetilde{\mathcal{Z}}$.  The dashed curve represents the image
under ${\bf{M}}$ of $\partial\mathcal{D}$ in the plane $\mathcal{Z}$.

\begin{figure}[H]
  \begin{center}
    \includegraphics[width=0.7\textwidth]{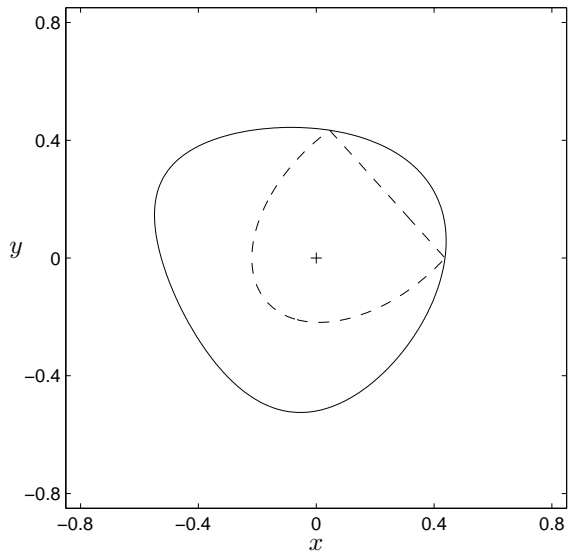}
    \caption{\label{fig:4} $\rho_1=\phi_5\phi_5^{\dagger}$ and
      $\rho_2=\phi_6\phi_6^{\dagger}$.  The origin $\rho_0$ is the
      maximally mixed state $I/3$.}
  \end{center}
\end{figure}

Figure~\ref{fig:4} shows a section with the maximally mixed state at
the origin, that is, $\rho_0=\widetilde{\rho}_0=I/3$.  We have chosen
the plane $\mathcal{Z}$ to go through the pure states
$\rho_1=\phi_5\phi_5^{\dagger}$ and $\rho_2=\phi_6\phi_6^{\dagger}$
corresponding to the arbitrarily chosen zeros number 5 and 6 of the
extremal witness $A$.  The section $\mathcal{Z}\cap\mathcal{D}$, shown
by the dashed curve, is of the type~C shown in
Figure~\ref{fig:illustrasjon}.  The section
$\widetilde{\mathcal{Z}}\cap\mathcal{D}$, shown by the full drawn
curve, is of type~A.  The pure states $\rho_1$ and $\rho_2$ are mapped
to rank two states $\widetilde{\rho}_1$ and $\widetilde{\rho}_2$, on
the boundary $\partial\mathcal{D}$.

\begin{figure}[H]
\begin{center}
  \includegraphics[width=0.7\textwidth]{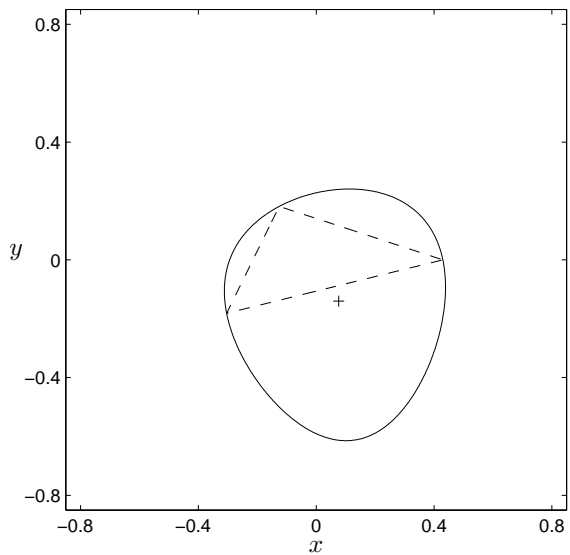}
  \caption{\label{fig:2} $\rho_1=\phi_5\phi_5^{\dagger}$ and
    $\rho_2=\phi_6\phi_6^{\dagger}$.  The origin $\rho_0$ is an even
    mix of the pure states $\rho_1$, $\rho_2$, and
    $\rho_3=\phi_7\phi_7^{\dagger}$.  The triangle is a face of an
    eight dimensional simplex defined by the nine zeros of the
    witness.}
\end{center}
\end{figure}

In Figure~\ref{fig:2} a different origin is used, we have chosen
$\rho_0$ as an even mix of the three pure states
$\rho_1=\phi_5\phi_5^{\dagger}$, $\rho_2=\phi_6\phi_6^{\dagger}$, and
$\rho_3=\phi_7\phi_7^{\dagger}$, which correspond to zeros of $A$ and
are mapped to rank two states $\widetilde{\rho}_1$,
$\widetilde{\rho}_2$, and $\widetilde{\rho}_3$.  We observe that the
projection of the maximally mixed state $I/3$ is off center.  This
section $\mathcal{Z}\cap\mathcal{D}$ is of the type~D shown in
Figure~\ref{fig:illustrasjon}, whereas the section
$\widetilde{\mathcal{Z}}\cap\mathcal{D}$ is still of type~A.
Figures 5-9 show various types of sections, as explained in the
figure captions.

\begin{figure}[H]
\begin{center}
  \includegraphics[width=0.7\textwidth]{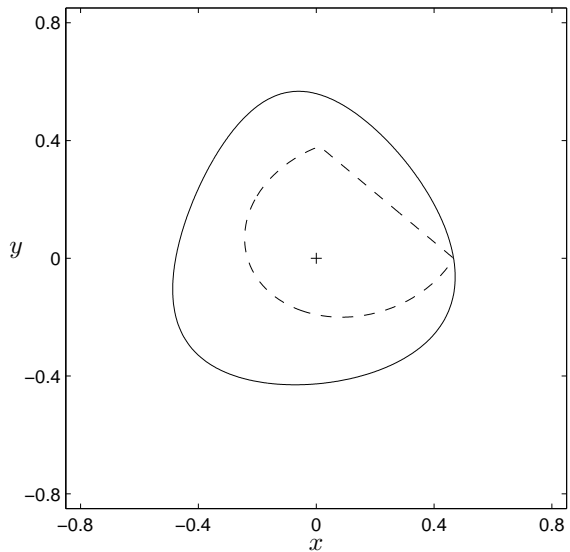}
  \caption{\label{fig:3} $\rho_1=\phi_1\phi_1^{\dagger}$, $\rho_2$ is
    a random pure state, and $\rho_0=I/3$.  Here
    $\widetilde{\rho}_1={\bf{M}}\rho_1$ has rank two and lies on the
    boundary $\partial\mathcal{D}$, whereas
    $\widetilde{\rho}_2={\bf{M}}\rho_2$ has full rank and lies in the
    interior of $\mathcal{D}$.  The section
    $\mathcal{Z}\cap\mathcal{D}$ is of type~C.}
\end{center}
\end{figure}

\begin{figure}[H]
\begin{center}
  \includegraphics[width=0.7\textwidth]{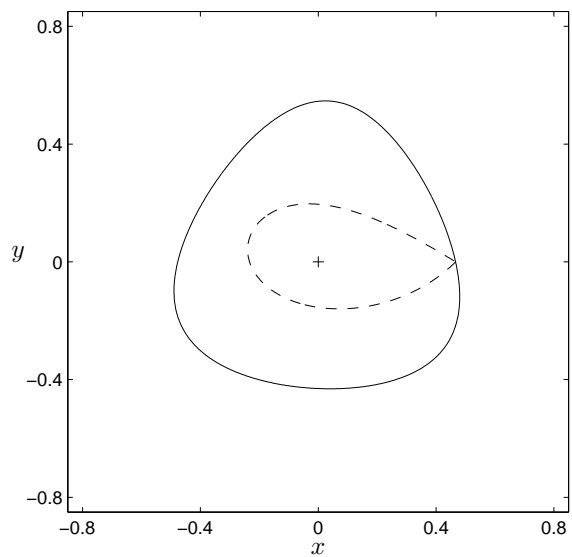}
  \caption{\label{fig:6} $\rho_1=\phi_1\phi_1^{\dagger}$, $\rho_2$ is
    a random state of rank three, and $\rho_0=I/3$.  Again $\rho_1$ is
    mapped to $\partial\mathcal{D}$ while $\rho_2$ is mapped to the
    interior of $\mathcal{D}$.  The section
    $\mathcal{Z}\cap\mathcal{D}$ is of type~B.}
\end{center}
\end{figure}

\begin{figure}[H]

\begin{center}
  \includegraphics[width=0.7\textwidth]{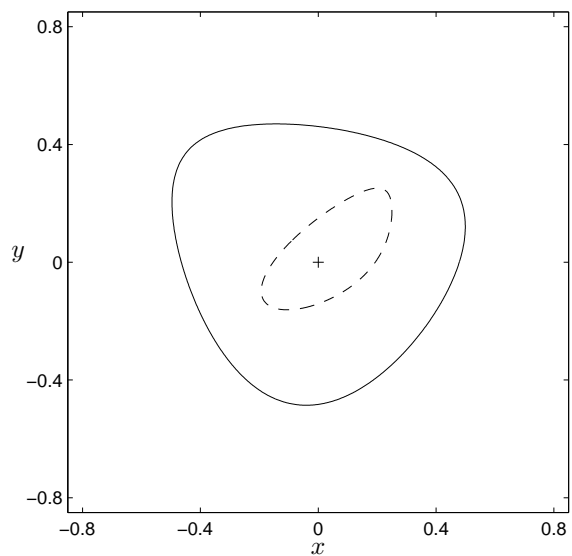}
  \caption{\label{fig:5} Both $\rho_1$ and $\rho_2$ are random states
    of rank three, and $\rho_0=I/3$.  Both sections in the plot are of
    type~A.}
\end{center}
\end{figure}

\begin{figure}[H]

\begin{center}
  \includegraphics[width=0.7\textwidth]{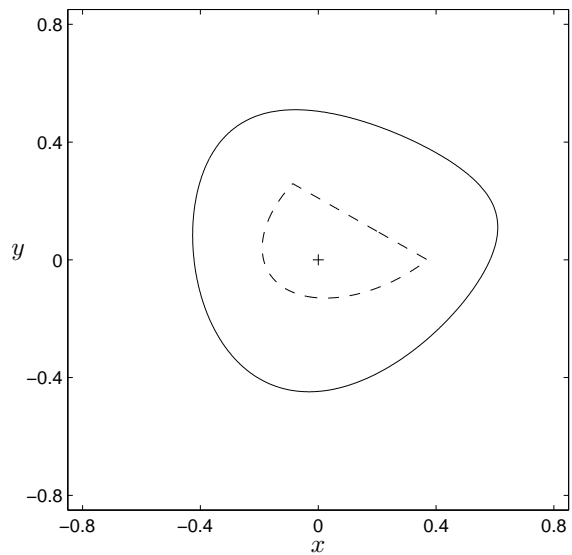}
  \caption{ Here both $\rho_1$ and $\rho_2$ are random pure states,
    and $\rho_0=I/3$.  The section $\mathcal{Z}\cap\mathcal{D}$ is of
    type~C, it is mapped entirely to the interior of $\mathcal{D}$.}
\end{center}
\end{figure}

\vspace*{-1\baselineskip}

\begin{figure}[H]

\begin{center}
  \includegraphics[width=0.7\textwidth]{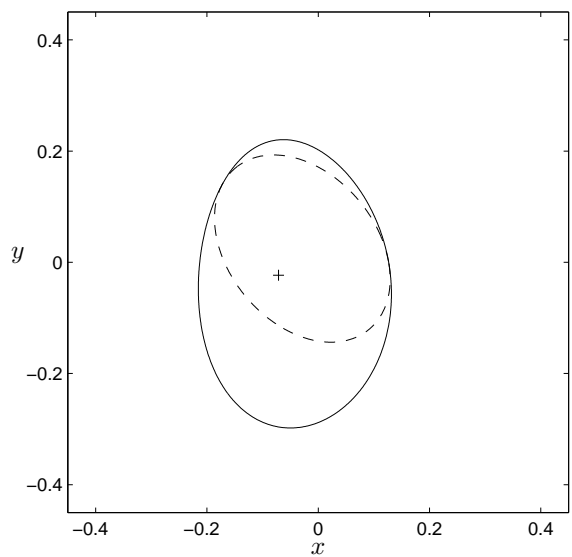}
  \caption{\label{fig:7} $\rho_1=\phi_1\phi_1^{\dagger}$ and
    $\rho_2=\phi_2\phi_2^{\dagger}$.  The origin $\rho_0$ is an even
    mix of $\rho_1$, $\rho_2$, and $\rho_3=\xi\xi^{\dagger}$ where
    $\xi$ is a linear combination of $\phi_1$ and $\phi_2$.  The
    section $\mathcal{Z}\cap\mathcal{D}$ is of type~E, it cuts through
    a Bloch sphere in $\partial\mathcal{D}$.  Its circular shape is
    distorted by the map into an ellipse.  The pure states $\rho_1$
    and $\rho_2$ are mapped to rank two states, on
    $\partial\mathcal{D}$, whereas all the other pure states on the
    surface of the Bloch sphere are mapped to rank three states.  The
    section $\widetilde{\mathcal{Z}}\cap\mathcal{D}$ is of type~A, its
    shape is neither circular nor elliptical.}
\end{center}
\end{figure}


\subsection{A second example: The Choi--Lam map}

The unital and trace preserving map
\begin{equation}
\label{eq:ChoiLammap}
  {\bf{M}}:X\mapsto
  Y=\frac{1}{2}
  \begin{pmatrix}
    X_{11}+X_{33} & -X_{12} & -X_{13}\\
    -X_{21} & X_{11}+X_{22} & -X_{23}\\
    -X_{31} & -X_{32} & X_{22}+X_{33}
  \end{pmatrix}
\end{equation}
was introduced by Choi and Lam in 1977 as the first example of an
extremal non-decomposable positive map~\cite{Choi1975,ChoiLam1977}. It
has been generalized to a one parameter family of extremal positive
maps, but it is still one of the very few known analytical examples of
such maps.  It corresponds to the extremal entanglement witness $W$
with

\settowidth{\mycolwd}{$-1$}
\begin{equation}
W^P=\left(
\begin{array}{*{9}{@{}C{\mycolwd}@{}}}
  1   & \cdot & \cdot & \cdot &  -1   & \cdot & \cdot & \cdot &  -1\\
  \cdot &   1   & \cdot & \cdot & \cdot & \cdot & \cdot & \cdot & \cdot\\
  \cdot & \cdot & \cdot & \cdot & \cdot & \cdot & \cdot & \cdot & \cdot\\
  \cdot & \cdot & \cdot & \cdot & \cdot & \cdot & \cdot & \cdot & \cdot\\
   -1   & \cdot & \cdot & \cdot &   1   & \cdot & \cdot & \cdot &  -1\\
  \cdot & \cdot & \cdot & \cdot & \cdot &   1   & \cdot & \cdot & \cdot\\
  \cdot & \cdot & \cdot & \cdot & \cdot & \cdot &   1   & \cdot & \cdot\\
  \cdot & \cdot & \cdot & \cdot & \cdot & \cdot & \cdot & \cdot & \cdot\\
   -1   & \cdot & \cdot & \cdot &  -1   & \cdot & \cdot & \cdot &   1
\end{array}
\right)
\end{equation}

where for clarity the many zero entries are represented as dots.  The
witness $W$ has three isolated quartic zeros,
\begin{equation}
	\label{Choipvdisc}
	e_{13}=e_1\otimes e_3\;,
	\qquad
	e_{21}=e_2\otimes e_1\;,
	\qquad
	e_{32}=e_3\otimes e_2\;,
\end{equation}
where $e_1,e_2,e_3$ are the natural basis vectors in
$\mathbb{C}^3$, and a continuum of quartic zeros $\phi\otimes\phi$
where
\begin{equation}
  \label{Choipvcont}
  \phi=\phi(\alpha,\beta)=
       e_1+\textrm{e}^{\textrm{i}\alpha}\,e_2
          +\textrm{e}^{\textrm{i}\beta}\,e_3\;,
\end{equation}
and $\alpha,\beta\in\mathbb{R}$.

Equation~\eqref{eq:ChoiLammap} has a very simple geometrical
interpretation.  In the subspace of diagonal matrices there is a
rotation by $60^{\circ}$ about the maximally mixed state, and in the
subspace of off-diagonal matrices there is an inversion.  There is
also an overall contraction by a factor of one half.

\begin{figure}[H]
\begin{center}
  \includegraphics[width=0.7\textwidth]{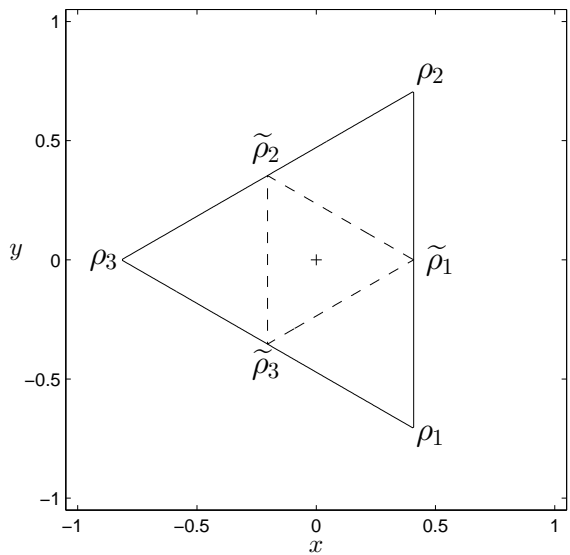}
  \caption{\label{fig:8} The Choi--Lam map in the plane of diagonal
    matrices.  $\rho_0=I/3$, $\rho_1=e_1e_1^{\dagger}$ and
    $\rho_2=e_2e_2^{\dagger}$.}
\end{center}
\end{figure}

Figure~\ref{fig:8} shows the special section through $\mathcal{D}$
containing the diagonal matrices, including the maximally mixed state
$\rho_0=I/3$ and the three pure states
\begin{equation}
  \rho_1=e_1e_1^{\dagger}=
  \begin{pmatrix}
    1 & 0 & 0\\
    0 & 0 & 0\\
    0 & 0 & 0
  \end{pmatrix},
  \quad
  \rho_2=e_2e_2^{\dagger}=
  \begin{pmatrix}
    0 & 0 & 0\\
    0 & 1 & 0\\
    0 & 0 & 0
  \end{pmatrix},
  \quad
  \rho_3=e_3e_3^{\dagger}=
  \begin{pmatrix}
    0 & 0 & 0\\
    0 & 0 & 0\\
    0 & 0 & 1
  \end{pmatrix},
\end{equation}
corresponding to the three isolated zeros of the witness $W$.  This
section is mapped into itself by the map ${\bf{M}}={\bf{M}}_W$, since
\begin{equation}
\widetilde{\rho}_1=(\rho_1+\rho_2)/2\;,\qquad
\widetilde{\rho}_2=(\rho_2+\rho_3)/2\;,\qquad
\widetilde{\rho}_3=(\rho_3+\rho_1)/2\;.
\end{equation}
Figure 11 shows a section defined
by the same pure states $\rho_1$ and $\rho_2$ as in Fig. 10, together
with a random state of rank three.  This section is not mapped into
itself.
%

\begin{figure}[H]

\begin{center}
  \includegraphics[width=0.7\textwidth]{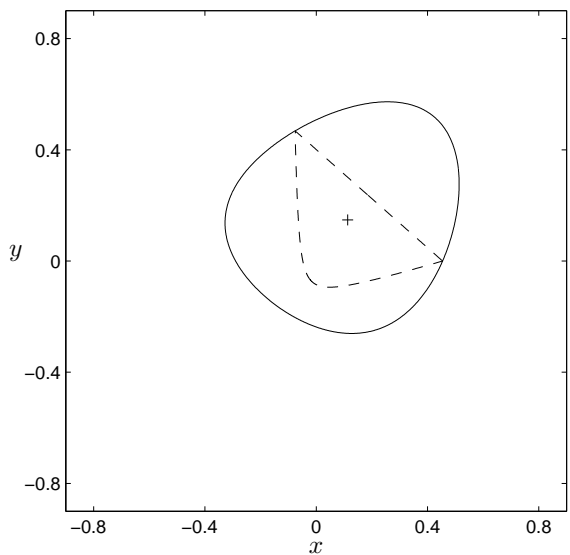}
  \caption{\label{fig:13} The Choi--Lam map.  Another section with
    $\rho_1=e_1e_1^{\dagger}$ and $\rho_2=e_2e_2^{\dagger}$.  The
    origin $\rho_0$ is chosen here as a random state of rank three, so
    the section $\mathcal{Z}\cap\mathcal{D}$ contains neither
    $e_3e_3^{\dagger}$ nor $I/3$.}
\end{center}

\end{figure}

\begin{figure}[H]
\begin{center}
  \includegraphics[width=0.7\textwidth]{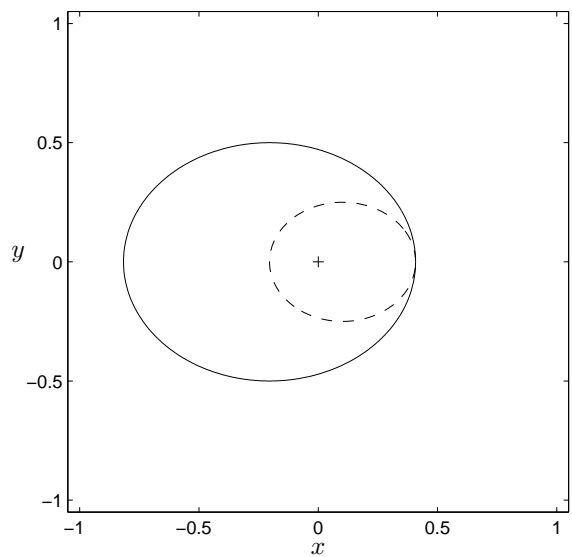}
  \caption{\label{fig:9} The Choi--Lam map.  $\rho_0=I/3$,
    $\rho_1=\rho(0,0)$, and $\rho_2=\rho_1+D(0,0)$, as defined in the
    text.}
\end{center}
\end{figure}

The continuous set of quartic zeros of the witness $W$, given in
Equation~\eqref{Choipvcont}, defines a two dimensional surface of pure
states
\begin{equation}
\label{eq:rhocont0}
\rho(\alpha,\beta)
=\frac{1}{3}\,
\phi(\alpha,\beta)(\phi(\alpha,\beta))^{\dagger}
=\rho_0+\sigma(\alpha,\beta)
\end{equation}
mapped by ${\bf{M}}$ to the boundary $\partial\mathcal{D}$.  Here
$\rho_0=I/3$, and $\sigma(\alpha,\beta)$ is a completely off-diagonal
matrix.  We see directly from Equation~\eqref{eq:ChoiLammap} that
\begin{equation}
{\bf{M}}\rho(\alpha,\beta)
=\rho_0-\frac{1}{2}\,\sigma(\alpha,\beta)
=\frac{1}{2}\,(3\rho_0-\rho(\alpha,\beta))\;.
\end{equation}
Thus the straight line through ${\bf{M}}\rho(\alpha,\beta)$ and
$\rho_0$ contains the pure state $\rho(\alpha,\beta)\,$.  This
surface of pure states is curved, but we may choose our plane
$\mathcal{Z}$ in such a way that it is tangent to the surface.  Define
for example
\begin{equation}
\xi(\alpha)=
\frac{\partial}{\partial\alpha}\,\phi(\alpha,\beta)
=\textrm{i\,e}^{\textrm{i}\alpha}\,e_2\;.
\end{equation}
Then the matrix
\begin{equation}
D(\alpha,\beta)=
\frac{\partial}{\partial\alpha}\,\rho(\alpha,\beta)
=\frac{1}{3}\left(
 \xi(\alpha)(\phi(\alpha,\beta))^{\dagger}
+\phi(\alpha,\beta)(\xi(\alpha))^{\dagger}\right)
\end{equation}
is a tangent to the surface such that
\begin{equation}
\rho(\alpha,\beta)+\epsilon D(\alpha,\beta)=
\rho(\alpha+\epsilon,\beta)+\mathcal{O}(\epsilon^2)\;.
\end{equation}
Figure~\ref{fig:9} shows a section where we have chosen $\rho_0=I/3$
and
\begin{eqnarray}
  \rho_1&=&\rho(0,0)
  =\frac{1}{3}\,(e_1+e_2+e_3)(e_1+e_2+e_3)^{\dagger}\:,
\\
  \rho_2&=&\rho_1+D(0,0)
  =\rho_1+
  \frac{\textrm{i}}{3}\,
  (e_2(e_1+e_3)^{\dagger}-(e_1+e_3)e_2^{\dagger})\;.
\end{eqnarray}
It does not matter that $\rho_2\not\in\mathcal{D}$.  The coordinate
axes as defined in Equation~\eqref{eq:ABaxes} are now
\begin{equation}
\label{eq:BCfig9}
B=a(\rho_1-\rho_0)
=\frac{a}{3}
  \begin{pmatrix}
    0 & 1 & 1\\
    1 & 0 & 1\\
    1 & 1 & 0
  \end{pmatrix},
\qquad
C=b(\rho_2-\rho_1)=bD(0,0)
=\frac{b}{3}
  \begin{pmatrix}
    0          &-\textrm{i} & 0\\
    \textrm{i} & 0          & \textrm{i}\\
    0          &-\textrm{i} & 0
  \end{pmatrix}.
\end{equation}
According to Equation~\eqref{eq:ChoiLammap} this section is mapped
into itself by a $180^{\circ}$ rotation and a scaling by 1/2,
\begin{equation}
B\mapsto\widetilde{B}=-\frac{1}{2}\,B\;,
\qquad
C\mapsto\widetilde{C}=-\frac{1}{2}\,C\;.
\end{equation}
Both sections $\mathcal{Z}\cap\mathcal{D}$ and
$\widetilde{\mathcal{Z}}\cap\mathcal{D}$ shown in Figure~\ref{fig:9}
are of type~F in our classification.  The numerical factors in
Equation~\eqref{eq:BCfig9} making $\widetilde{B}$ and $\widetilde{C}$
orthonormal are $a=\sqrt{6}\,$, $b=3\,$.

%


%
\begin{figure}[H]
\begin{center}
  \includegraphics[width=0.7\textwidth]{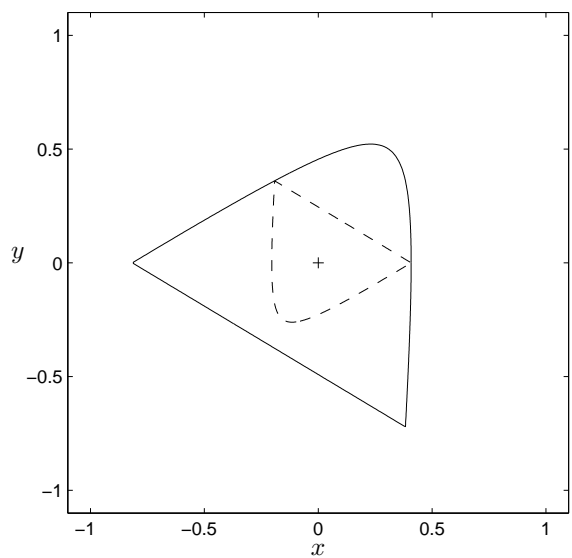}
  \caption{\label{fig:10} The Choi--Lam map.  We use here $\rho_0=I/3$
    and two randomly chosen pure states
    $\rho_1=\phi_1\phi_1^{\dagger}$, $\rho_2=\phi_2\phi_2^{\dagger}$
    corresponding to product vectors $\phi_1\otimes\phi_1^{\ast}$,
    $\phi_2\otimes\phi_2^{\ast}$ from the continuum of quartic zeros.}
\end{center}
\end{figure}
%


%
\begin{figure}[H]

\begin{center}
  \includegraphics[width=0.7\textwidth]{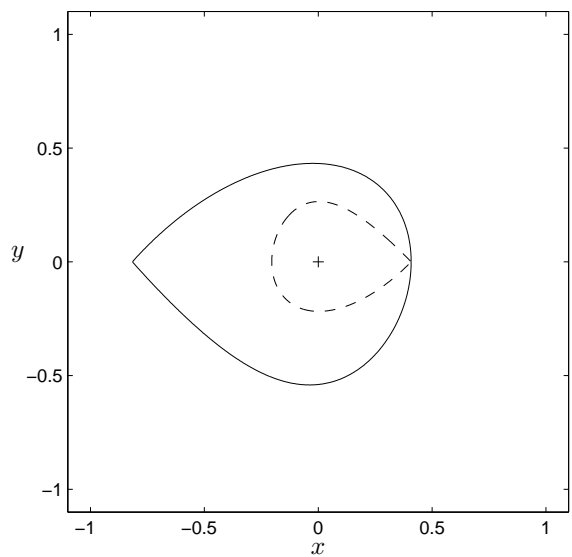}
  \caption{\label{fig:11} The Choi--Lam map.  $\rho_0=I/3$,
    $\rho_1=e_1e_1^{\dagger}$, and $\rho_2$ is a random state of rank
    three.}
\end{center}

\end{figure}

\begin{figure}[H]

\begin{center}
  \includegraphics[width=0.7\textwidth]{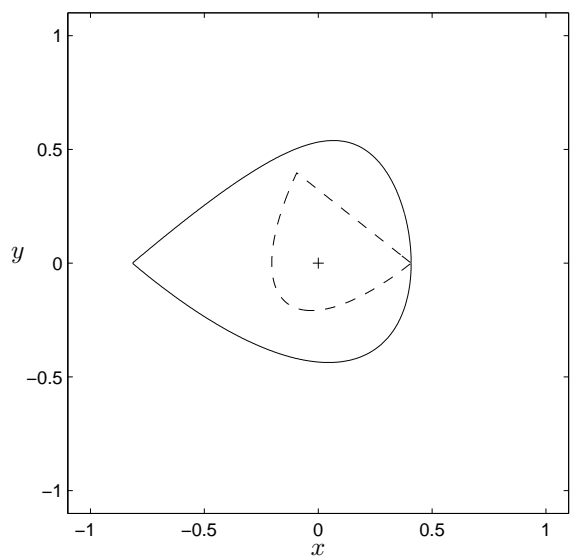}
  \caption{\label{fig:12} The Choi--Lam map.  $\rho_0=I/3$,
    $\rho_1=e_1e_1^{\dagger}$, and $\rho_2$ is a random pure state.}
\end{center}

\end{figure}

Figures 13-15 illustrate the Choi-Lam map by other
sections, as explained in the figure captions.

\subsection{A third example in $2\times 4$ dimensions}
\label{sec:Athirdexample}

Our third example is based on the study of optimal witnesses by
Lewenstein et al.~\cite{Lewenstein2000}.  As an example they describe
how to create optimal witnesses for proving the entanglement of the
PPT states in $2\times 4$ dimensions discovered by Pawe{\l}
Horodecki~\cite{Horodecki1997}.  We have chosen arbitrarily the
parameter value $b=0.6$ and computed numerically the optimal witness
as described in~\cite{Lewenstein2000}.  It has exactly eight zeros,
all quadratic.  It is not extremal, but is the centre of a two
dimensional circular face of the set of normalized witnesses
(normalized to have unit trace).  The boundary of this face is a
circle of quartic extremal witnesses, one of which defines the map
described numerically in the Appendix.  Note that this map is neither
unital nor trace preserving.

This extremal witness has two separate continuous rings of zeros,
which are all necessarily quartic, since they are not discrete.  To a
zero $\phi\otimes\chi$ with $\phi^{\dagger}\phi=\chi^{\dagger}\chi=1$
corresponds a density matrix
\begin{equation}
\phi\phi^{\dagger}=\frac{1}{2}
\begin{pmatrix}
1+z & x-\textrm{i}y\\
x+\textrm{i}y & 1-z
\end{pmatrix}
\end{equation}
with $x^2+y^2+z^2=1$.  This rank one density matrix in $\mathcal{D}_2$
is mapped to a rank three density matrix in $\mathcal{D}_4$ with
$\chi$ as an eigenvector of eigenvalue zero.  Thus the image of the
two rings on the boundary of $\mathcal{D}_2$ is where the image of
$\mathcal{D}_2$ touches the inside of the boundary of $\mathcal{D}_4$.

The coordinates $x,y,z$ defining a zero are given by one parameter
$\theta$ as follows.  Let
\begin{equation}
a =  0.1807362587783353\;,\quad
b =  0.047422228589395\;,\quad
\theta_0 = 1.121090508802759\;,
\end{equation}
\begin{equation}
s = \frac{a\cos(2\theta+\theta_0)}{\cos(\theta-\theta_0)}\;,\qquad
t = \frac{-bs\pm\sqrt{1+s^2-b^2}}{1+s^2}\;,
\end{equation}
\begin{equation}
x = t\cos\theta\;,\qquad
y = t\sin\theta\;,\qquad
z = -b-ts\;.
\end{equation}
Note that $t\to 0$ and $ts\to -b\pm 1$ as $s\to\pm\infty$.

All the extremal quartic witnesses forming the circular boundary of
the face have similar sets of zeros, it is only the parameter
$\theta_0$ that varies as we go around the boundary.  Any two of them
have exactly eight zeros in common, corresponding to the special
values $\theta=0$ and $\theta=\pm\pi/3$, where $x,y,z$ are independent
of $\theta_0$, and to $\theta=\theta_0+\pi/2$, where $x=y=0$,
$z=\pm 1$.  Every witness in the interior of the face has exactly
these eight zeros.

The Figures~\ref{fig:2x4a} to~\ref{fig:2x4c} show the zeros of the
witness defining the map given in the Appendix.  The
Figures~\ref{fig:2x4d} to~\ref{fig:2x4f} show the same zeros, and in
addition the zeros of two other extremal witnesses on the boundary of
the same face.  The zeros of different such extremal witnesses are
found simply by changing the value of $\theta_0$.

\begin{figure}[H]
\begin{center}
  \includegraphics[width=0.65\textwidth]{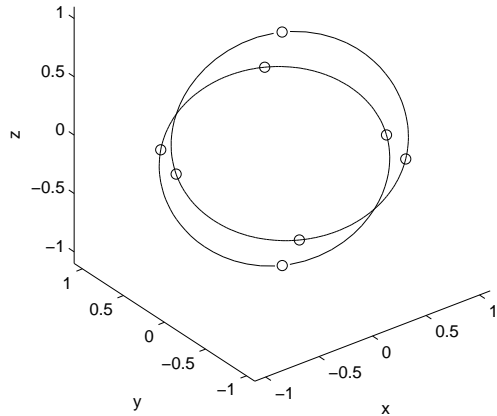}
  \caption{\label{fig:2x4a} The two rings of zeros of the extremal
    witness in dimension $2\times 4$, on the surface of the unit
    sphere.  The eight zeros of the optimal witness are marked by
    small circles.}
\end{center}
\end{figure}

\begin{figure}[H]
\begin{center}
  \includegraphics[width=0.7\textwidth]{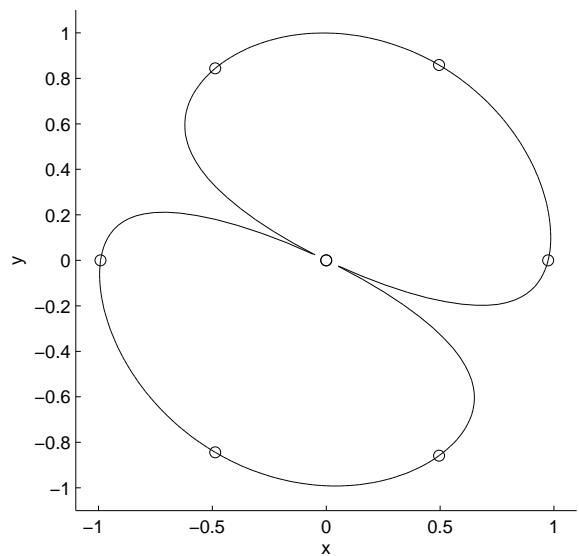}
  \caption{\label{fig:2x4b} The $xy$ projection of
    Figure~\ref{fig:2x4a}.  Two of the eight special zeros are at the
    origin $x=y=0$, the six others are at angles $n\pi/3$ from the $x$
    axis with $n=0,1,2,3,4,5$.}
\end{center}
\end{figure}

\begin{figure}[H]
\begin{center}
  \includegraphics[width=0.7\textwidth]{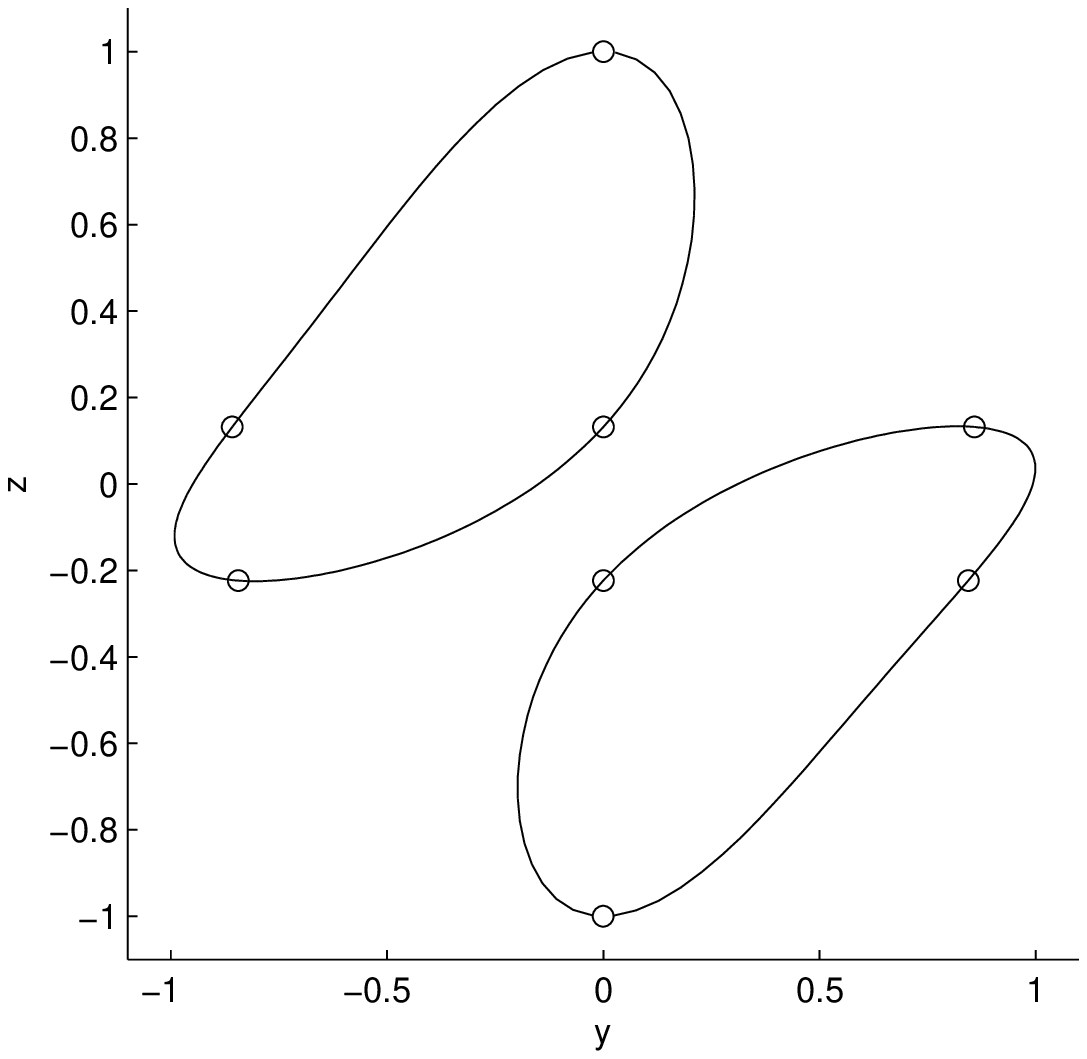}
  \caption{\label{fig:2x4c} The $yz$ projection of
    Figure~\ref{fig:2x4a}.}
\end{center}
\end{figure}

\begin{figure}[H]
\begin{center}
  \includegraphics[width=0.7\textwidth]{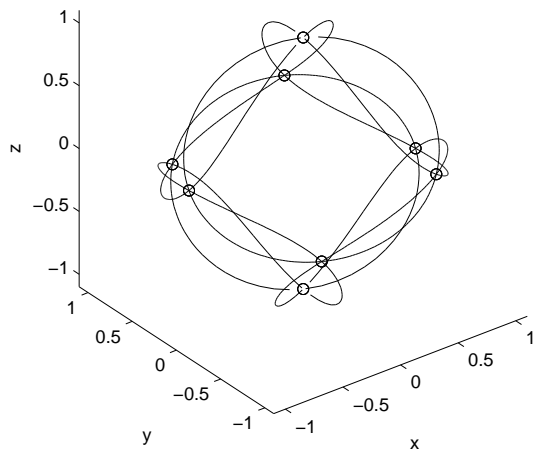}
  \caption{\label{fig:2x4d} The same as Figure~\ref{fig:2x4a}, but
    including the zeros of two other extremal witnesses with different
    values of $\theta_0$.}
\end{center}
\end{figure}

\begin{figure}[H]
\begin{center}
  \includegraphics[width=0.7\textwidth]{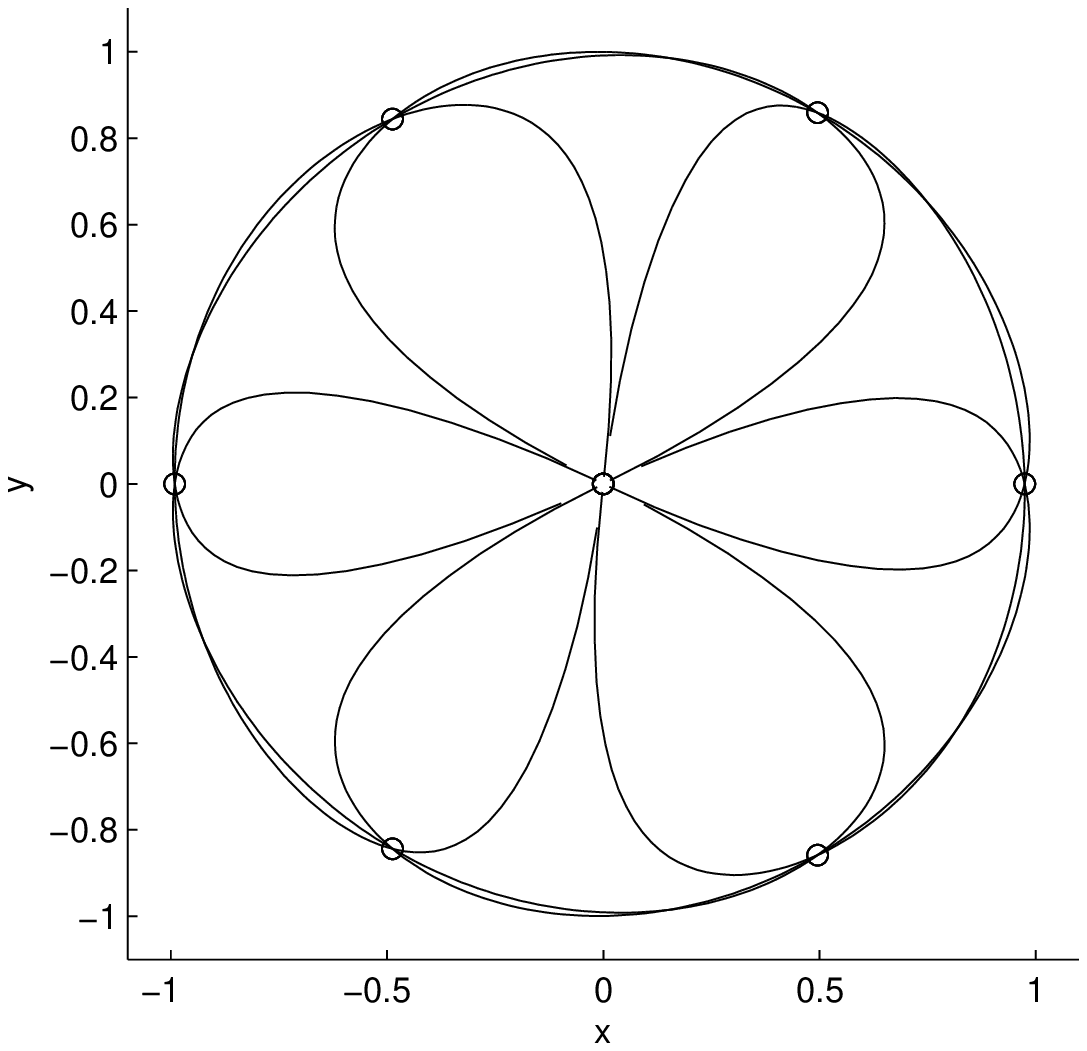}
  \caption{\label{fig:2x4e} The $xy$ projection of
    Figure~\ref{fig:2x4d}.}
\end{center}
\end{figure}

\begin{figure}[H]
\begin{center}
  \includegraphics[width=0.7\textwidth]{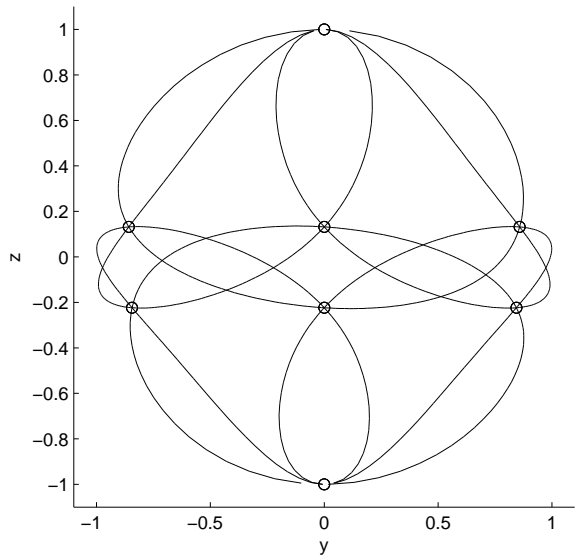}
  \caption{\label{fig:2x4f} The $yz$ projection of
    Figure~\ref{fig:2x4d}.}
\end{center}
\end{figure}


\section{Summary and conclusions}

The purpose of the work presented here has been to gain a more
intuitive understanding of the geometry of positive maps, which are
related to entanglement witnesses in bipartite quantum systems.  We
try to visualize the action of an extremal positive map by plotting
various two dimensional sections through the set of density matrices.

For this purpose it is useful to transform the map to some standard
form.  We argue that any positive map can be transformed into a unital
and trace preserving form through a product transformation of the
corresponding entanglement witness.  If the witness lies in the
interior of ${\mathcal{S}^{\circ}}$, which means that it has no zeros,
this follows from the proof of a similar result given
in~\cite{Leinaas2006}.  We present an iteration scheme for computing
the transformation numerically, and we find in practice in $3\times 3$
dimensions that it works well even for extremal witnesses and other
witnesses lying on the boundary of ${\mathcal{S}^{\circ}}$.  We find
numerically that the unital and trace preserving form of a positive
map is unique up to unitary product transformations.

We present plots related to two different extremal entanglement
witnesses in $3\times 3$ dimensions and one in $2\times 4$ dimensions.
The first example is a randomly chosen generic extremal entanglement
witness with quadratic zeros found in previous numerical searches.  We
produce the corresponding positive map
$\mathbf{M}:H_3\rightarrow H_3$, and then plot two dimensional
sections in order to illustrate how the image
$\mathbf{M}\mathcal{D}_3$ lies inside $\mathcal{D}_3$.

We then repeat this scheme for a version of the Choi--Lam map, or
entanglement witness, which is extremal but highly non-generic, having
only quartic zeros.  It has three isolated quartic zeros, and one
continuous two dimensional set of zeros which are necessarily quartic.

The most important feature of the plots is related to the zeros of the
two witnesses.  A zero defines a pure state in $\mathcal{D}_3$ which
is mapped to the boundary of $\mathcal{D}_3$.  In particular, a
generic extremal witness in $3\times 3$ dimensions, like the one
presented here, has nine zeros, defining a simplex in $\mathcal{D}_3$
with nine vertices which is mapped to a simplex in $\mathcal{D}_3$
with nine vertices touching the boundary of $\mathcal{D}_3$ from the
inside.

We should emphasize that we have studied here only $3\times 3$ and
$2\times 4$ dimensions, which are the simplest nontrivial cases.  In
higher dimensions the symmetric dimensions $m\times n$ with $m=n>3$
are clearly the most interesting.  The complexity increases much with
the dimension, because the simplex of pure states in $\mathcal{D}_3$
corresponding to the quadratic zeros of the extremal witness
considered here, becomes a polytope with a number of vertices larger
than $m^2$ when $m>3$.  For example, with $m=n=4$ the minimum number
of zeros of a quadratic extremal witness is 20, as compared to the
dimension of $\mathcal{D}_4$ which is 15.  With $m=n=5$ the minimum
number of zeros is 37, as compared to the dimension of $\mathcal{D}_5$
which is 24~\cite{Hansen2014}.

We believe that the geometrical way of thinking illustrated here may
be a fruitful approach when one wants to construct examples of
extremal maps and entanglement witnesses.  It may be that the increase
in complexity with increasing dimension, which is a well known
phenomenon, is easier to handle geometrically than by other methods.

\section*{Acknowledgments}
  We acknowledge gratefully a research grant from The Norwegian
  University of Science and Technology (Leif Ove Hansen).  We thank
  Erling St{\o}rmer for his interest in our studies of extremal
  positive maps, and in particular his question what they look like,
  which inspired the present paper.

\appendix

\section{An extremal positive map from $H_2$ to $H_4$}

We introduce the following $4\times 4$ matrices.
\begin{equation}
B_0+B_3=
\begin{pmatrix}
a_1 &-\textrm{i}a_3 & \textrm{i}a_4 & a_1\\
\textrm{i}a_3 & a_2 & 0 &\textrm{i}a_3\\
-\textrm{i}a_4 & 0 & a_2 &-\textrm{i}a_4\\
a_1 &-\textrm{i}a_3 & \textrm{i}a_4 & a_1
\end{pmatrix}
\end{equation}
with
%
\begin{eqnarray}
a_1&=&0.0244482760740412\;,\quad
a_2=0.2152770862261020\;,\nonumber\\
a_3&=&0.0114377547217477\;,\quad
a_4=0.0500075452822933\;.
\end{eqnarray}
\begin{equation}
B_0-B_3=
\begin{pmatrix}
a_5 & \textrm{i}a_7 & \textrm{i}a_8 &-a_5\\
-\textrm{i}a_7 & a_6 & 0 & \textrm{i}a_7\\
-\textrm{i}a_8 & 0 & a_6 & \textrm{i}a_8\\
-a_5&-\textrm{i}a_7 &-\textrm{i}a_8 & a_5
\end{pmatrix}
\end{equation}
with
\begin{eqnarray}
a_5&=&0.0644909685779951\;,\quad
a_6=0.1957836691218616\;,\nonumber\\
a_7&=&0.0774551312933996\;,\quad
a_8=0.0177155824920755\;.
\end{eqnarray}
\begin{equation}
B_1=
\begin{pmatrix}
0 &-a_9-\textrm{i}a_{10} & a_{11}-\textrm{i}a_{12} &-\textrm{i}a_{13}\\
-a_9+\textrm{i}a_{10} & 0 &-a_{14}-\textrm{i}a_{15} & a_{11}+\textrm{i}a_{16}\\
a_{11}+\textrm{i}a_{12} &-a_{14}+\textrm{i}a_{15} & 0 &-a_9-\textrm{i}a_{17}\\
\textrm{i}a_{13} & a_{11}-\textrm{i}a_{16} &-a_9+\textrm{i}a_{17} & 0
\end{pmatrix}
\end{equation}
with
\begin{eqnarray}
\!\!\!\!\!
\!\!\!\!\!
a_9&=&0.0363521121932822\;,\quad
a_{10}=0.0276760626964089\;,\quad
a_{11}=0.0094553411157518\;,\nonumber\\
\!\!\!\!\!
\!\!\!\!\!
a_{12}&=&0.0293657267910500\;,\quad
a_{13}=0.0130745578191192\;,\quad
a_{14}=0.1714859526438769\;,\\
\!\!\!\!\!
\!\!\!\!\!
a_{15}&=&0.0675990471881839\;,\quad
a_{16}=0.0121590711417975\;,\quad
a_{17}=0.0384768416753617\;.\nonumber
\end{eqnarray}
\begin{equation}
B_2=
\begin{pmatrix}
0 &-a_{11}-\textrm{i}a_{12} &-a_{9}+\textrm{i}a_{10} & \textrm{i}a_{18}\\
-a_{11}+\textrm{i}a_{12} &-a_{14} & \textrm{i}a_{19} & a_{9}-\textrm{i}a_{17}\\
-a_{9}-\textrm{i}a_{10} &-\textrm{i}a_{19} & a_{14} & a_{11}-\textrm{i}a_{16}\\
-\textrm{i}a_{18} &-a_{9}+\textrm{i}a_{17} & a_{11}+\textrm{i}a_{16} & 0
\end{pmatrix}
\end{equation}
with
\begin{equation}
a_{18}=0.0082070224528484\;,\quad
a_{19}=0.0424325553291989\;.
\end{equation}
The $2\times 2$ matrix
\begin{equation}
A=\frac{1}{2}
\begin{pmatrix}
u+z & x-\textrm{i}y\\
x+\textrm{i}y & u-z
\end{pmatrix}
\end{equation}
is positive when $u>0$ and $u^2\geq x^2+y^2+z^2$.  In
Section~\ref{sec:Athirdexample} we discuss the following extremal
positive map, which we do not transform to unital and trace preserving
form,
\begin{equation}
{\bf{M}} : A\mapsto B=uB_0+xB_1+yB_2+zB_3\;.
\end{equation}


\begin{thebibliography}{99}

\bibitem{Schrodinger}
E.~Schr\"{o}dinger,\\
\emph{Discussion of probability relations between separated systems}\\
Proc.~Camb.~Philos.~Soc.~\textbf{31}, 555 (1935)

\bibitem{EPR1935}
A.~Einstein, B.~Podolsky, and N.~Rosen,\\
\emph{Can Quantum-Mechanical Description of Physical Reality Be Considered Complete?}\\
Phys.~Rev. \textbf{47}, 777 (1935)

\bibitem{Nielsen}
M.~Nielsen and I.~Chuang,\\
\emph{Quantum Computation and Quantum Information}\\
Cambridge University Press, Cambridge (2000)

\bibitem{MPRHorodecki96}
M.~Horodecki, P.~Horodecki, and R.~Horodecki,\\
\emph{Separability of mixed states: necessary and sufficient conditions}\\
Phys.~Lett.~A \textbf{223}, 1 (1996)

\bibitem{Barbieri2003}
M.~Barbieri, F.~De Martini, G.~Di Nepi, P.~Mataloni, G.M.~D'Ariano, and\\ C.~Macchiavello,\\
\emph{Detection of Entanglement with Polarized Photons: Experimental Realization of an\\ Entanglement Witness}\\
Phys.~Rev.~Lett. \textbf{91}, 227901 (2003)

\bibitem{Jamiolkowski1972}
A.~Jamio{\l}kowski,\\
\emph{Linear transformations which preserve trace and positive semidefiniteness of operators}\\
Rep.~Math.~Soc.~\textrm{3}, 275 (1972)

\bibitem{Choi1975}
M.D.~Choi,\\
\emph{Positive semidefinite biquadratic forms}\\
Linear Algebra Appl.~\textbf{12}, 95 (1975)

\bibitem{Stormer1963}
E.~St\o rmer,\\
\emph{Positive linear maps of operator algebras}\\
Acta Math.~\textbf{110}, 233 (1963)

\bibitem{Stormer2013}
E.~St\o rmer,\\
\emph{Positive linear maps of operator algebras}\\
Monographs in Mathematics, Springer Verlag, Berlin (2013)

\bibitem{Peres1996}
A.~Peres,\\
\emph{Separability criterion for density matrices}\\
Phys.~Rev.~Lett.~\textbf{77}, 1413 (1996)

\bibitem{Hansen2014}
L.O.~Hansen, A.~Hauge, J.~Myrheim, and P.\O.~Sollied,\\
\emph{Extremal entanglement witnesses}\\
arXiv:1305.2385

\bibitem{Lewenstein2000}
M.~Lewenstein, B.~Kraus, J.~I.~Cirac, and P.~Horodecki,\\
\emph{Optimization of entanglement witnesses}\\
Phys.~Rev.~A~\textbf{62}, 052310 (2000)

\bibitem{Stormer1982}
E.~St\o rmer,\\
\emph{Decomposable linear maps on $C^{*}$-algebras}\\
Proc.~Amer.~Math.~Soc.~\textrm{86}, 402 (1982)

\bibitem{Woronowicz1976}
S.L.~Woronowicz,\\
\emph{Positive maps of low dimensional matrix algebras}\\
Rep.~Math.~Phys.~\textrm{10}, 165 (1976)

\bibitem{Leinaas2006}
J.M.~Leinaas, J.~Myrheim, and E.~Ovrum,\\
\emph{Geometrical aspects of entanglement}\\
Phys.~Rev.~A \textrm{74}, 012313 (2006)

\bibitem{ChoiLam1977}
M.D.~Choi and T.Y.~Lam,\\
\emph{Extremal positive semidefinite forms}\\
Math.~Ann.~\textbf{231}, 1 (1977)

\bibitem{Horodecki1997}
P.~Horodecki,\\
\emph{Separability criterion and inseparable mixed states
with positive partial transposition}\\
Phys.~Lett.~A~\textbf{232}, 333 (1997)





\end{thebibliography}
\end{document}